\documentstyle[prb,aps,eqsecnum,multicol,ifthen]{revtex}
\newcommand{\showfigures}{yes}     
\newcommand{\wide}[2]{
\end{multicols}
\widetext
\noindent
\ifthenelse{\equal{#1}{t}}
{}
{
\raisebox{0.1in}[0in][0.02in]{$\rule{3.575in}{0.002in}
\rule{0.002in}{0.08in}$}
}
#2
\ifthenelse{\equal{#1}{b}}
{}
{
{\raisebox{-0.1in}[0in][0.02in]
{\hspace{3.575in}$\rule{0.002in}{0.08in}
\rule[0.08in]{3.575in}{0.002in}$}
}
}
\begin{multicols}{2}
\noindent
}
\tighten
\ifthenelse{\equal{\showfigures}{yes}}{\input psfig}{} 
\begin{document}
\draft
\title{ Divergence of the Classical Trajectories and Weak Localization}
\author{ I.L. Aleiner$^{1,*}$ and A.I. Larkin$^{1,2}$}  
\address{$^1$Theoretical Physics Institute, University of Minnesota, 
 Minneapolis, MN 55455 \\
$^2$L.D. Landau Institute for Theoretical Physics, 117940 Moscow, Russia}
\date{Submitted to Phys. Rev. B, March 15, 1996; Revised 
version submitted on July 16, 1996.}
\maketitle
\begin{abstract}
 We study the weak localization correction (WLC) to
transport coefficients    of a  system of electrons in a static
long-range potential (e.g. an antidot array or   ballistic cavity). We
found that the weak localization correction to the current response is
delayed by the large time $t_E = \lambda^{-1} |\ln
\hbar|$, where
$\lambda$ is the Lyapunov exponent. In the semiclassical regime $t_E$ is
much larger than the transport lifetime. Thus, the fundamental
characteristic of the classical chaotic motion, Lyapunov exponent, may be
found by measuring the frequency or temperature dependence of WLC.
\end{abstract}

\begin{multicols}{2} 

\section{Introduction}

An electron system in a static potential is characterized by the
following  linear scales: the geometrical size of the system, $L$;  the
transport mean free path
$l_{tr}=v_F\tau_{tr}$ being the characteristic distance at which a
particle can travel before the direction of its momentum is randomized; 
the characteristic scale  the potential energy changes over, $a$; and 
de~Broglie wavelength $\lambda_F$, (for the Fermi system
$\lambda_F= \hbar/p_F$, with $p_F=mv_F$ being the Fermi momentum). In the
most important metallic regime
$\lambda_F\ll L, l_{tr}$.  The  scale of the potential $a$ may be
arbitrary and  depending upon this scale two regimes can be
distinguished:\\
\indent i) Quantum chaos (QC), $a^2 > \lambda_Fl_{tr}$.\\
\indent ii) Quantum disorder (QD), $a^2 < \lambda_Fl_{tr}$.\\ The physics
behind this distinction is quite transparent: after an electron 
interacts with the scatterer of the size
$a$, the quantum uncertainty in the direction of  its momentum
$\delta\theta$  is of the order of $\delta\theta\simeq
\lambda_F/a$. Therefore, the uncertainty in the position of the particle
$\delta x$ on the next scatterer can be estimated as $\delta x \simeq
l_{tr}\delta\theta \simeq
\l_{tr} \lambda_F/a$.\cite{footnote1} If $\delta x \ll a$, the quantum
uncertainty in the position of the particle is not important and its
motion can be described by the classical  Hamilton (or Liouville)
equations. Except some special cases, these equations are not integrable,
the electron trajectory is extremely sensitive to the initial conditions
and the classical motion is chaotic. The quantum phenomena in such regime
still bear essential features of the classical motion; it is accepted in
the literature to call such regime ``quantum chaos''. In the opposite
limit,  
$\delta x \gg a$ and the electron looses any memory about its classical
trajectory already after the first scattering. Any disordered system where
the Born approximation is applicable may serve as an example of QD regime.

Under assumption of the ergodicity of the system, the  classical
correlator is usually found from the Boltzmann or diffusion equations. 
The form of these equations  is identical for both regimes. The only
difference appears in the expression for the cross-section   entering
into the collision integral. For the QC, this cross-section can be found
by solving  the classical equations of motion, whereas in  the QD it is
determined by solution of the corresponding quantum mechanical scattering
problem.

Subject of weak localization (WL) theory is the study of the first order
in
$\lambda_F/l_{tr}$ corrections to the transport coefficients of the
system. 
 The WL in the quantum disorder have been studied for more than fifteen
years already\cite{g4,Gorkov,review}.  
 The regime of the quantum chaos attracted attention only
recently\cite{Klitzing,Markus,Chang,Stone,Mello,Argaman}. This interest
was motivated mostly by   technological advances which allowed the
fabrication of the structures where
$a \gg \lambda_F$. Two examples of these structures are: (1) the antidot
arrays\cite{Klitzing} where role of $a$ is played by the diameter of an
antidot; (2) ballistic cavities\cite{Markus,Chang} where $a \simeq l_{tr}
\simeq L$ coincides with the size of the cavity.

Weak localization corrections are known to have anomalous dependence upon
the frequency $\omega$, temperature  or the applied magnetic field and
that is why they can be experimentally observed. For the two-dimensional
system case $L \to\infty$ the WL correction to the conductivity
$\Delta\sigma$ can be conveniently written as
\begin{equation}
\Delta\sigma =-
\frac{e^2s}{4\pi^2\hbar}\Gamma(\omega)
\ln\left(\frac{1}{\omega\tau_{tr}}\right),\quad \omega\tau_{tr} \lesssim
1,
\label{renfunction}
\end{equation} where $s=2$ is the spin degeneracy, and $\Gamma(\omega)$
is  a renormalization function. It is this function where the difference
between the quantum disorder and quantum chaos is drastic. Gorkov, Larkin
and Khmelnitskii\cite{Gorkov} showed  that, for the whole frequency
domain, $\Gamma = 1$ for the quantum disorder and does not depend upon
the details of the scattering. The question is: Does  such a universality
persist for the quantum chaos too? 

In this paper, we will show that, in the limit $\omega \to 0$, the
renormalization function $\Gamma \to 1$ which proves the universality of
weak localization correction for the quantum chaos\cite{interaction}.
However, unlike for the quantum disorder, $\Gamma $ acquires the frequency
dependence at $\omega$ much smaller than $1/\tau_{tr}$. This frequency
dependence  can be related to the Lyapunov exponent $\lambda$
characterizing the classical motion of the particle. It gives an
opportunity to extract the value of the Lyapunov exponent
 from the measurements of the frequency dependence of the conductivity.
We found
\begin{equation}
\Gamma (\omega) = \exp\left(2i\omega t_E -\frac{ 2\omega^2 \lambda_2
t_E}{\lambda^2}\right),
\label{result}
\end{equation} where Ehrenfest time $t_E$ is  the time it takes for the
minimal wave packet to spread over the distance of the order of $a$ and
it is given by\cite{footnote1}
\begin{equation} t_E = \frac{1}{\lambda}\ln
\left(\frac{a}{\lambda_F}\right).
\label{tE}
\end{equation} Quantity
$\lambda_2\simeq\lambda$  in Eq.~(\ref{result}) characterizes the
deviation of the Lyapunov exponents, and it will be explained in
Sec.~\ref{sec:2} in more details. In the time representation, result
(\ref{result}) corresponds to the delay of the weak localization
correction to the current response by large time $2t_E$, see
Fig.~\ref{fig:0}.

{\narrowtext 

\begin{figure}[h]
\vspace{-7mm}
\ifthenelse{\equal{\showfigures}{yes}}
{\hspace*{0.5cm}\psfig{figure=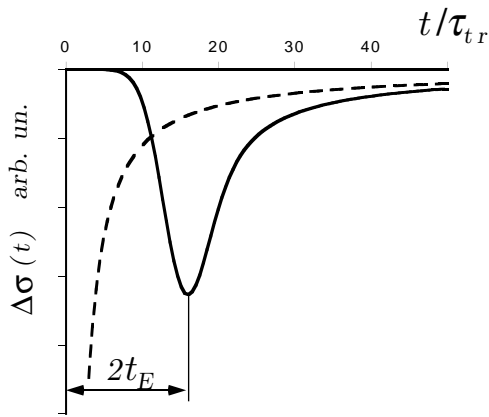,height=3.6in}} {\vspace*{3.6in}}
\vspace{-3.0cm}
\caption{The weak localization correction to the conductivity in the time
domain,
$\Delta\sigma (t) =
\int
\frac{d \omega}{2\pi}\Delta\sigma(\omega)e^{-i\omega t}$ for the quantum
chaos (solid line) and quantum  disorder (dashed line) regimes. The
developed theory is valid for $t \protect\gtrsim t_E$. Solid curve is
calculated for parameters $\lambda=4\lambda_2=1/\tau_{tr},\ \ln
\left(a/\lambda_F\right) =7$.}
\label{fig:0}
\end{figure} }

The paper is arranged as follows. In Sec.~\ref{sec:2}, we present the
phenomenological derivation of  Eq.~(\ref{result}).  The explicit
expression relating the weak localization correction to the  solution of
the Liouville equation will be derived in Sec.~\ref{sec:3}.  In
Sec.~\ref{sec:4}, we will find the quantum corrections to the
conductivity in the infinite chaotic system. Sec.~\ref{sec:5}  describes
the effects of the magnetic field and finite phase relaxation time on the
renormalization function. The  conductance of the ballistic cavities is
studied in Sec.~\ref{sec:6}. Our findings are summarized in Conclusion.

\section{Qualitative discussion}
\label{sec:2}

Classical diffusion equation is based on the assumption that at long time
scales  an electron looses any memory about its previous experience.
However, during its travel, the electron may traverse  the  same spatial
region and be affected by the same scatterer more than once. These two
scattering events are usually considered independently, because with the
dominant probability the electron enters this region having completely
different momentum.

However, if we wish to find the probability $W_0(T,\rho_0)$ for a particle
to have the momentum opposite to the initial one, ${\bf p}(T)=-{\bf
p}(0)$, (time
$T$ is much larger than $\tau_{tr}$) and to approach its starting point
at small distance
$|{\bf r}(T)-{\bf r}(0)|=\rho_0
\ll a$,  we should take into account the fact that the motion of the
particle at the initial and final stages are correlated. This is because
the trajectory along which   the particle moves on the final stage,
$\left[{\bf r}(T-t), {\bf p}(T-t)\right]$ almost coincides with the
trajectory particle moved along at the initial stage, $\left[{\bf r}(t),
{\bf p}(t)\right]$,  see Fig.~\ref{fig:1}.  These correlations break down
the description of this problem by the diffusion equation. The behavior of
the distribution function for this case can be related to the Lyapunov
exponent, and we now turn to the discussion of such a relation. 
 (Relevance of
$W_0(T,\rho_0)$ to the weak localization correction will become clear
shortly.)

{\narrowtext 

\begin{figure}[h]
\ifthenelse{\equal{\showfigures}{yes}}
{\hspace*{0.5cm}\psfig{figure=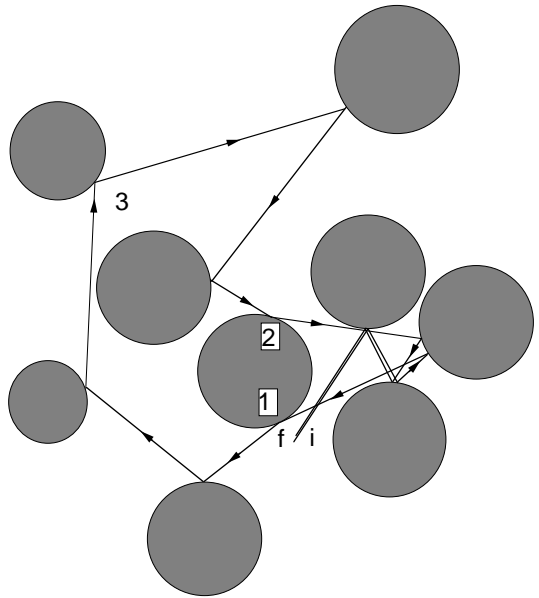,height=3.1in}} {\vspace*{3.1in}}
\vspace{0.4cm}
\caption{The classical trajectory corresponding to the probability of
return at the initial point with the momentum opposite to the initial
one. In the ``Lyapunov region'' the initial ``i-1'' and final ``2-f''
fragments  of the trajectory are governed by the same potential.}
\label{fig:1}
\end{figure} }

The correlation of the motion at the final  and initial stages can be
conveniently characterized by two  functions
\begin{equation}
\mbox{\boldmath $\rho$}(t)={\bf r}(t)-\!{\bf r}(T-t), \ \ {\bf k}(t)={\bf
p}(T-t)+{\bf p}(t).
\label{rhok}
\end{equation} The classical equations of motion for these functions are
\begin{mathletters}
\begin{eqnarray} &\displaystyle{\frac{
                      \partial \mbox{\boldmath $\rho$}
                     }{\partial t} =\frac{{\bf k}(t)}{m},}&
\label{eqm1}\\ &\displaystyle{\frac
                    {\partial {\bf k}}
                    {\partial t} =\frac{\partial U[{\bf r}(T-t)]}
            {\partial {\bf r}} -
\frac{\partial U[{\bf r}(t)]}
        {\partial {\bf r}} }&,
\label{eqm2}
\end{eqnarray}
\label{eqmotion} where $U$ is the potential energy. If the distance
$\rho$ is much larger than the characteristic spatial scale of the
potential $a$, Eqs.~(\ref{eqmotion}) lead to the usual result
$\langle\rho(t)\rangle \propto t^{1/2}$ at times $t$ much larger than
$\tau_{tr}$. Situation is different, however, for $\rho \ll a$, where the
diffusion equation is not applicable (we will call this region of the
phase space the ``Lyapunov region''). Thus, the calculation of function
$W_0(T,\rho_0)$ should be performed in two steps. First, we have to
calculate the conditional  probability $W(a,\rho_0;t)$, which is defined
so that the probability  for the distance $\rho(t)$  to become larger
than   $a$ during the time interval $[t,\ t+\Delta t]$ is equal to
$W(a,\rho_0;t)\Delta t$ under the condition  $\rho (0) =\rho_0$. Second,
we have to obtain the probability
$W_D(a,t)$ for the diffusively moving particle to approach its starting
point to the distance of the order of $a$ (it corresponds to the fragment
``1-3-2'' in Fig.~\ref{fig:1}). Then, the function
$W_0(T,\rho_0)$ is given by
\end{mathletters}
\begin{equation} W_0(T,\rho_0) =\int_0^{T}dtW_D(a,T-2t)W(a,\rho_0;t). 
\label{integral}
\end{equation}

Now, we perform the first step: finding of the probability
$W(a,\rho_0;t)$. We consider more general quantity $W(\rho,\rho_0;t)$ for
$\rho < a$.  We expand the right hand side of Eq.~(\ref{eqm2}) up to the
first order in
$\rho$ which yields
\begin{equation}
\frac
                    {\partial {k}_j}
                    {\partial t} =-{\cal M}_{ij}(t)\rho_i,\quad {\cal
M}_{ij}(t)\equiv\frac{\partial^2 U[{\bf r}(t)]}
            {\partial {r}_i\partial {r}_j}.
\label{expanded}
\end{equation} It is easily seen from Eq.~(\ref{expanded}) and
Fig.~\ref{fig:1}, that the change in the momentum $k$  during  the
scattering event is proportional to the distance $\rho$. On the other
hand, it follows from Eq.~(\ref{eqm1}) that the change in the value of
$\rho$ between scattering events is proportional to $k$. Therefore, one
can expect that the distance $\rho$ grows exponentially with time. In
Appendix~{\ref{ap:1} we  explicitly solve  the model of weak dilute
scatterers $l_{tr} \gg a$ and find the expression for the distribution
function $W(\rho) =\left<\delta(t - t(\rho))\right>$, where 
$\left<...\right>$ means average over directions of ${\bf p}$.  Here we
present  qualitative arguments which enable us to establish the form of
the function $W$ for the general case.

We notice that, if matrix
$\hat{\cal M}(t)$ does not depend on time, the solution of
Eqs.~(\ref{eqm1}) and (\ref{expanded}) is readily available:
\begin{equation}
\rho (t) \simeq \rho (0)e^{\lambda t},
\label{constant}
\end{equation} where the quantity $\lambda$ is related to the maximal
negative eigenvalue of $\hat{\cal M}$. We will loosely call $\lambda$ the
Lyapunov exponent. If $\hat{\cal M}$ varies with time,  the solution of
Eqs.~(\ref{eqm1}) and (\ref{expanded}) is not possible. We argue,
however, that for the large time $t\gg \tau_{tr}$, this variation may be
described by a random correction to the Lyapunov exponent:
\begin{equation}
\frac{d\ln \rho }{dt} =\lambda + \delta\lambda(t).
\label{Lognorm}
\end{equation}  At time scale larger than $\tau_{tr}$ the correlation
between the values of $\delta\lambda(t)$ at different moments of time can
be neglected,
 $\left<\delta\lambda(t_1)\delta\lambda(t_2)\right> =\lambda_2
\delta(t_1-t_2)$, that immediately gives the log-normal form for the
function $W$:
\begin{eqnarray} &&\displaystyle{W(\rho,\rho_0; t) = 
            \sqrt{\frac{\lambda^3}
                       {2\pi\lambda_2\cal L(\rho)}
                      }
\exp\left[-\frac{
                 \lambda\left(
                      {\cal L(\rho)}
                       -
                 \lambda t\right)^2
                  }
 {2\lambda_2{\cal L(\rho)}}\right],} \nonumber\\
&&\displaystyle{\hspace{4cm} {\cal L(\rho)} =\ln
                  \rho/\rho_0.}
\label{Lognormresult}
\end{eqnarray} Formula (\ref{Lognormresult}) is valid in general case
even though analytic calculation of the values of $\lambda$ and
$\lambda_2$ (as well as of the diffusion constant) can be performed only 
for some special cases, {\em e.g.} for $l_{tr} \gg a$.  For the antidot
arrays, $\lambda$ is given by the inverse scattering time up to the
factor of the order of $\ln (l_{tr}/a)$.\cite{Krylov} The model of the
dilute weak scatterers is considered in Appendix~\ref{ap:1}. The result is
$\lambda,\lambda_2\simeq\tau_{tr}^{-1}(l_{tr}/a)^{2/3}$.  In the ballistic
billiards, coefficients
$\lambda,\lambda_2$ are of the order of the inverse flying time across the
system.

Equation~(\ref{Lognormresult}) describes the distribution function only in
the vicinity of its maximum,
$\left|\ln\left({\rho}/{\rho_0}\right)-\lambda t
\right|
\lesssim \lambda t$. However, this result will be sufficient if time $T$
in Eq.~(\ref{integral}) is large enough $T\gtrsim {\cal L}(a)/2\lambda$.
At smaller times the probability of return is determined by the tail of
the distribution function $W(\rho)$ which is by no means log-normal.

It is worth mentioning, that there is some arbitrariness in our choice of
the initial conditions ${\bf p}(T)=-{\bf p}(0)$,
$|{\bf r}(T)-{\bf r}(0)|=\rho_0$. The other possible choice is $|{\bf
p}(T)+{\bf p}(0)| =k_0$, $|{\bf r}(T)-{\bf r}(0)|=0$. In this case,
formula (\ref{Lognormresult}) remains valid upon the substitution $\rho_0
\to ak_0/p(0)$.

Now we can find $W_0(T,\rho_0)$ from Eq.~(\ref{integral}).  Substituting
Eq.~(\ref{Lognormresult})  into Eq.~(\ref{integral}), we arrive to the 
result for the probability $W_0(T,\rho_0)$:
\begin{eqnarray}
&&\displaystyle{W_0(T,\rho_0)=\int\frac{d\omega}{2\pi}W_0(\omega,\rho_0)e^{-i\omega
T}}, \label{probability}\\
&&\displaystyle{W_0(\omega,\rho_0)\!=\!W_D(\omega,a)\!
\exp\left(\frac{2i\omega{\cal L}(a)}{\lambda}\! - \!
\frac{2
\omega^2 \lambda_2 {\cal L}(a)} {\lambda^3}\!\right),}
\nonumber
\end{eqnarray} where $W_D(\omega,a)$ is the Fourier transform of the
function  $W_D(t,a)$. 
  Function $W_D(a,\omega)=W_D(\omega; a \to l_{tr})W_D(\omega,l_{tr}) $ is
determined by two consecutive processes. First process, with the
probability
$W_D(\omega; a \to l_{tr})$, is the separation of the trajectories from
distance $a$, at which they become independent to the distance larger than
$l_{tr}$, where the diffusion equation is applicable. The characteristic
time for such process is of the order of
$\tau_{tr}$, and thus $W_D(\omega; a \to l_{tr}) =1 +{\cal
O}(\omega\tau_{tr})$. The probability $W_D(\omega,l_{tr})$ is found by
solving the standard diffusion equation. For the two-dimensional case,
which will be most interesting for us, function $W_D(\omega,a)$ has the
form
\begin{equation} W_D(\omega,a) = \frac{1}{4\pi
D}\ln\left(\frac{1}{\omega\tau_{tr}}\right),
\label{diffuson}
\end{equation}  where $D=v_F^2\tau_{tr}/2$ is the diffusion constant.
Notice that this function does not depend on $a$. Expressions
(\ref{diffuson}) and (\ref{Lognormresult}) are written with the
logarithmic accuracy.

So far, we considered a purely classical problem. We found the probability
for a particle, propagating in a classical disordered potential,  to
approach its starting point with the momentum opposite to its initial
one. In the calculation of the classical kinetic coefficients (e.g.
conductivity), the integration over all the direction of the momentum is
performed. As the result, the peculiarities in the probability discussed
above are washed out and do not appear in the classical kinetic
coefficients.  However, the function $W_0(\rho,t)$ plays very important
role in the semiclassical approach to some quantum mechanical problems. 
One of such problems  arose long time ago in the study\cite{Larkin68} of
break down of  the method of the quasiclassical trajectories in the
superconductivity theory\cite{diGennesShapoval}.  Another problem is the
weak localization  in the quantum chaos and we turn to the study of this 
phenomenon  now.

It is well
known\cite{LarkinKhmelnitskii,AronovAltshuler,ChakravartySchmidt} that
the probability $w$ for the particle to get from, say, point $i$ to point
$f$, see Fig.~\ref{fig:2}a, can be obtained by, first, finding the
quasiclassical amplitudes $A_{\alpha}$ for different paths connecting the
points, and, then, by squaring the modulus of their sum:
\begin{equation} w=\left|\sum_{\alpha}A_\alpha\right|^2=
\sum_{\alpha}\left|A_\alpha\right|^2 +
\sum_{\alpha\neq\beta}A_\alpha A_\beta^\ast.
\label{qprobability}
\end{equation}  The first term in Eq.~(\ref{qprobability}) is nothing
else but the sum of the classical probabilities of the different paths,
and the second term is due to the quantum mechanical interference of the
different amplitudes.  For generic pairs $\alpha,\beta$, the product
$A_\alpha A_\beta^\ast$ oscillates strongly on the scale of the order of
$\lambda_F$  as the function of the position of point $f$. This is because
the lengths of the paths
$\alpha$ and
$\beta$ are substantially different. Because all the measurable quantities
are averaged on the scale much larger than $\lambda_F$, such oscillating
contributions can be neglected. There are pairs of paths, however, which
are coherent. The example of such paths is shown in Fig.~\ref{fig:2}b,
(paths $1$ and $2$).  These paths almost always coincide.
 The only difference is that fragment $BEB$ is traversed in the opposite
directions by trajectories $1$ and $2$.  In the absence of the magnetic
field and the spin-orbit interaction, the phases of the amplitudes
$A_1$ and $A_2$ are equal because the lengths of the trajectories are
close. The region, where the distance between trajectories $1,2$ is
largest, see inset in Fig.~\ref{fig:2}b, deserves some discussion. At
this point the directions of the paths at points $B_1, B_2$ are almost
opposite to those at points
$B_1^\prime, B_2^\prime$. Furthermore, the differences between lengths of
 paths $1$ and $2$ should not be larger than $\lambda_F$. It imposes
certain restriction on angle $\delta\phi$ at which trajectory $2$ can
intersect itself and  on distance $\delta \rho$ to which the trajectory
$1$ can approach itself. Simple geometric consideration, self-evident from
inset in Fig.~\ref{fig:2}b, gives the estimate $\delta \rho \simeq
\sqrt{\lambda_Fl_{tr}}$ and  $\delta \phi \simeq \sqrt{\lambda_F/l_{tr}}$,
so that the uncertainty relation $\delta \phi\delta \rho \simeq \lambda_F$
holds. In other words, one of the trajectories  should almost ``graze
itself'' at the point
${B}$. 

{\narrowtext 
\begin{figure}[h]
\ifthenelse{\equal{\showfigures}{yes}}
{\hspace*{4mm}\psfig{figure=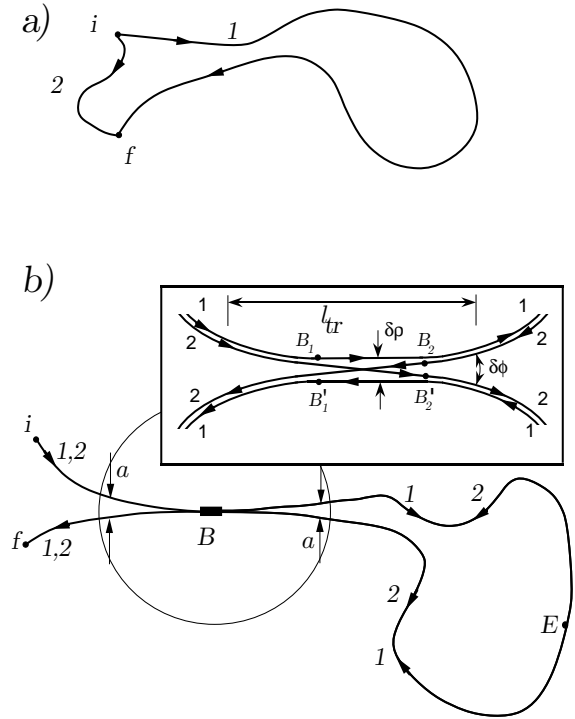,height=3.5in}} {\vspace*{3.5in}}
\vspace{0.8cm}
\caption{Examples of the classical (a) non-coherent and (b) coherent
paths between points $i$ and
$f$. The scatterers are not shown and the paths are straighten for
clarity. Encircled is the Lyapunov region. The region of the quantum
switch between trajectories (marked by the rectangular) is blown up on
the inset.}
\label{fig:2}
\end{figure} }

The interference part of the contribution of the coherent pairs to the
probability
$w$, see Eq.~(\ref{qprobability}) is of the same order as the classical
probability for these trajectories. Therefore, the contribution of the
interference effect to the conductivity  $\sigma$ is proportional to the
probability to find the trajectories similar to those from
Fig.~\ref{fig:2}b. In order to calculate this probability, we use
function $W_0(\rho,t)$ defined in the beginning of this section: the
probability $dP$ for a trajectory to graze itself during the time
interval $\left[ t_1, t_1+dt_1\right]$ is
\begin{eqnarray} dP_1=\delta
\rho\delta\phi v_Fdt_1W_0(\sqrt{\lambda_Fl_{tr}},t_1)=\label{dp1} \\
\lambda_Fv_Fdt_1W_0(\sqrt{\lambda_Fl_{tr}},t_1).
\nonumber
\end{eqnarray} in two dimensions. We are, however, interested in the
correction to the transport coefficients (such as the diffusion constant
or the conductivity). These quantities are contributed mostly by the
points $i,f$ located at the distance $\simeq l_{tr}$ from each other.
Thus, in order to contribute to the diffusion constant or the
conductivity, ends of the trajectories should separate
 from each other to the distance of the order of $a$, i.e. the
trajectories should overcome the Lyapunov region one more time. The
conditional probability $dP_2$ that the trajectories diverge at the
distance
$\sim a$  during the time interval $\left[ t_2, t_2+dt_2\right]$ under the
condition that the self grazing occurred at moment $t_1$ is given by
\begin{equation} dP_2=dt_2W(a,\sqrt{\lambda_Fl_{tr}},t_2-t_1).
\label{dp2}
\end{equation} where $W$ is given by Eq.~(\ref{Lognormresult}).

Summing over all the time intervals, we obtain for the quantum correction
to the conductivity
$\Delta\sigma $:
\begin{eqnarray} &&\hspace*{2cm}\frac{\Delta\sigma}{\sigma}\approx - \int
dP_1dP_2
\approx\label{corr1}\\  
&&v_F\lambda_F\!\!\int_0^\infty\!\!\!\! dt_2
\int_0^\infty\!\!\!\! dt_1 W_0\!\left(\sqrt{\lambda_Fl_{tr}},t_1)\right)
W\!\left(a,\sqrt{\lambda_Fl_{tr}},t_2\right).
\nonumber
\end{eqnarray} If the correction at finite frequency $\omega$ is needed,
the time integration in Eq.~(\ref{corr1}) should be replaced with the
Fourier transform over the total time of travel between points initial
and final points $t=2t_2+t_1$ in Eq.~(\ref{corr1}). This yields
\begin{equation} {\Delta\sigma(\omega)}=
-\frac{\sigma}{\pi\hbar\nu}W\left(a,\sqrt{\lambda_Fl_{tr}},2\omega\right)
W_0\left(\sqrt{\lambda_Fl_{tr}},\omega\right),
\label{corr2}
\end{equation} where $\nu$ is the density of states per one spin.
Coefficient in Eq.~(\ref{corr2}) and sign in Eqs.~(\ref{corr1}) and
(\ref{corr2}), known for the quantum disorder, will be reproduced for the
quantum chaos in Sec.~\ref{sec:3}. Substituting
Eqs.~(\ref{Lognormresult}) and (\ref{diffuson}) into Eq.~(\ref{corr2})
and using the Einstein relation
${\sigma} = se^2\nu D$, we arrive to the final result
(\ref{result}).\cite{footnote2}

\section{Weak localization in the quantum chaos}
\label{sec:3} It follows from  the previous discussion that the
calculation of the quantum correction is related to the  probability to
find a classical trajectory with large correlated segments. Standard
diagrammatic technique\cite{Gorkov,review,AronovAltshuler} is not
convenient for this case because the averaging over the disorder
potential is performed on the early stage, and including the additional
correlations is technically difficult. That is why we will derive the
expression for the quantum correction in terms of classical
probabilities, which are the solutions of the Liouville equation in a
given potential.  This result is important on its own, because it 
provides a tool for the description of the quantum effects in the 
ballistic cavities. The averaging, then, can be performed only on the
final stage of the calculations. For the sake of concreteness, we consider
two-dimensional case; generalization to the other dimensions is
straightforward. We will omit the Planck constant in all the intermediate
calculations.

\subsection{Introduction of basic quantities}

It is well known that transport coefficients  can be calculated using the
product of two exact Green functions $K_{\epsilon}$:
\begin{equation} K_{\epsilon}(\omega; {\bf r}_1, {\bf r}_2, {\bf r}_3,
{\bf r}_4) =G_{\epsilon+\frac{\omega}{2}}^R({\bf r}_1, {\bf r}_2)
G_{\epsilon-\frac{\omega}{2}}^A({\bf r}_3, {\bf r}_4).
\label{K}
\end{equation} Here $G^{R(A)}$ is the exact retarded (advanced) Green
function of the electron in the disordered potential $U({\bf r})$ and it
satisfies the equation
\begin{equation}
\left[\epsilon \pm i0 - \hat{H}_1\right]G_\epsilon^{R,A}({\bf r}_1,  {\bf
r}_2) = \delta({\bf r}_1-{\bf r}_2),
\label{G}
\end{equation} where one-electron Hamiltonian is given by
\begin{equation}
\hat{H}_1 = -\frac{\nabla_{{\bf}_1}^2}{2m} + U({\bf r_1}).
\label{Hamiltonian}
\end{equation}
 
For instance, the Kubo formula for the conductivity is
\begin{eqnarray*} &&\displaystyle{\sigma^{\alpha\beta}
\left(\omega;{\bf r}_1,{\bf r}_2\right)=\frac{se^2}{4m^2}\int
\frac{d\epsilon}{2\pi}
\left(-\frac{\partial f}{\partial\epsilon}\right)\times}  \\
&&\displaystyle{
\left[\nabla_{{\bf r}_1}^\alpha - \nabla_{{\bf r}_4}^\alpha\right]
\left[\nabla_{{\bf r}_3}^\beta - \nabla_{{\bf r}_2}^\beta\right]
K_{\epsilon}(\omega; {\bf r}_1, {\bf r}_2, {\bf r}_3, {\bf r}_4)\left|_{
\begin{array}{ll} {\bf r}_4={\bf r}_1\\ {\bf r}_3={\bf r}_2
\end{array}}\right. ,}
\nonumber
\end{eqnarray*} the expression for the polarization operator is
\begin{eqnarray}
\Pi
\left(\omega;{\bf r}_1,{\bf r}_2\right)&=&s\left[
\raisebox{0in}[3mm][3mm]{$\nu$}
\delta\left({\bf r}_1-{\bf r}_2\right)
 -\right.\nonumber\\
 &i&\!\!\omega\int\left.
\displaystyle{\frac{d\epsilon}{2\pi}
\frac{\partial f}{\partial\epsilon}} K_{\epsilon}(\omega; {\bf r}_1, {\bf
r}_2, {\bf r}_2, {\bf r}_1)
\right],
\label{polarization}
\end{eqnarray} and so on. Here $f(\epsilon)=\left(e^{(\epsilon-\mu
)/T}+1\right)^{-1}$ is the Fermi distribution function. Unfortunately,
the exact calculation of $K$ is not possible and  one has to resort on
some approximations. 

In general, function $K_{\epsilon}(\omega; {\bf r}_1, {\bf r}_2, {\bf
r}_3, {\bf r}_4)$ oscillates rapidly  with the distance between its
arguments. It contains  non-oscillating part only if its arguments are
paired: ${\bf r}_1={\bf r}_4,\ {\bf r}_2={\bf r}_3$ or, alternatively,
${\bf r}_1={\bf r}_3,\ {\bf r}_2={\bf r}_4$. If  they are not paired but 
still close to each other pairwise, then, it is very convenient to perform
the Fourier transform over the difference of these close arguments:
\begin{eqnarray} &&\displaystyle{K_{\epsilon}(\omega; {\bf r}_1, {\bf
r}_2,
 {\bf r}_3, {\bf r}_4)
 = \!\int\!\frac{d{\bf p}_1}{(2\pi)^2}\frac{d{\bf p}_2}{(2\pi)^2}
e^{i{\bf p}_1\left({\bf r}_1-{\bf r}_4\right)} e^{i{\bf p}_2\left({\bf
r}_3-{\bf r}_2\right)}}
\nonumber\\  &&\displaystyle{\quad \times K_{\epsilon}^{\cal
D}\left(\omega;\ {\bf p}_1, {\bf R}_1;
\ {\bf p}_2, {\bf R}_2\right),}\nonumber\\ &&\displaystyle{\quad\quad
{\bf R}_1=\frac{{\bf r}_1 +{\bf r}_4}{2},\ {\bf R}_2=\frac{{\bf r}_2+{\bf
r}_3}{2} }.
\label{mixedrepD}
\end{eqnarray} or, alternatively, 
\begin{eqnarray} &&\displaystyle{K_{\epsilon}(\omega; {\bf r}_1, {\bf
r}_2, {\bf r}_3, {\bf r}_4)
 = \!\int\!\frac{d{\bf p}_1}{(2\pi)^2}\frac{d{\bf p}_2}{(2\pi)^2}
e^{i{\bf p}_1\left({\bf r}_1-{\bf r}_3\right)} e^{i{\bf p}_2\left({\bf
r}_4-{\bf r}_2\right)}}
\nonumber\\  &&\displaystyle{\quad \times K_{\epsilon}^{\cal
C}\left(\omega;\ {\bf p}_1, {\bf R}_1;
\ {\bf p}_2, {\bf R}_2\right),}\nonumber\\ &&\displaystyle{\quad\quad
{\bf R}_1=\frac{{\bf r}_1 +{\bf r}_3}{2},\ {\bf R}_2=\frac{{\bf r}_2+{\bf
r}_4}{2} }.
\label{mixedrepC}
\end{eqnarray}

Let us now derive the semiclassical equation for the function $K^{\cal
D}$. From Eq.~(\ref{G}) and definition (\ref{K}) we can write the equation
for function $K$ in the form
\begin{eqnarray}
\left[\omega - \hat{H}_1 + \hat{H}_4\right] K_{\epsilon}(\omega; {\bf
r}_1, {\bf r}_2, {\bf r}_3, {\bf r}_4)
 = \nonumber\\ G_{\epsilon-\frac{\omega}{2}}^A({\bf r}_3, {\bf
r}_4)\delta({\bf r}_1-{\bf r}_2) -G_{\epsilon+\frac{\omega}{2}}^R({\bf
r}_1, {\bf r}_2)\delta({\bf r}_3-{\bf r}_4).
\label{eqK}
\end{eqnarray} If the distance $|r_1- r_4|$ is much smaller than the
characteristic scale
 of the potential, we expand term $\hat{H}_4 - \hat{H}_1$ in
Eq.~(\ref{eqK})  in distance $|r_1- r_4|$, and perform the Fourier
transform analogous to Eq.~(\ref{mixedrepD}).  The result can be 
expressed in terms   of the Liouvillean operator $\hat L$:
\begin{equation} i\left(\hat{H}_1 -
\hat{H}_4\right) \approx\hat{L}_1 =\frac{\partial {\cal H}}{\partial {\bf
p}_1}\cdot\frac{\partial}{\partial {\bf R}_1}-
\frac{\partial {\cal H}}{\partial {\bf R}_1}
\cdot\frac{\partial}{\partial {\bf p}_1}.
\label{Lioville}
\end{equation} where ${\cal H}({\bf p}, {\bf r})$ is the Hamiltonian
function
\begin{equation} {\cal H}({\bf p}, {\bf r}) = \frac{{\bf p}^2}{2m} +
U({\bf r}).
\label{HamiltonFunction}
\end{equation} With the help of  Eqs.~(\ref{eqK}), (\ref{Lioville}) and
(\ref{mixedrepD}), we  obtain
\begin{eqnarray}
\left[-i\omega + \hat{L}_1\right]K_{\epsilon}^{\cal D}\left(\omega;\ {\bf
p}_1, {\bf R}_1;
\ {\bf p}_2, {\bf R}_2\right) = \label{KD}\\ 
2\pi\delta\left[\epsilon-{\cal H}({\bf p}_2, {\bf R}_2)\right]
(2\pi)^2\delta({\bf p}_1-{\bf p}_2)
\delta({\bf R}_1-{\bf R}_2).
\nonumber
\end{eqnarray} Delta-functions in the right hand side of Eq.~(\ref{KD})
should be understood in a sense of there subsequent convolution  with a
function smooth on a spatial scale larger than $\lambda_F$. When deriving
Eq.~(\ref{KD}), we used the semiclassical approximation for the Green
functions  
\begin{equation} G_{\epsilon}^{R,A}({\bf r}_1, {\bf r}_2)\! = \!
\int\!\frac{d{\bf p}}{(2\pi)^2}
\frac{e^{i{\bf p} \left({\bf r}_1-{\bf r}_2\right)}}{\epsilon -{\cal
H}\left[{\bf p}, \left({\bf r}_1+{\bf r}_2\right)/2\right]\pm i0}
\label{semGreen}
\end{equation} in the right-hand side of Eq.~(\ref{eqK}) and neglected
small frequency $\omega$ in comparison with the  large energy $\epsilon
\simeq E_F$.

Liouvillean operator (\ref{Lioville}) describes the motion of an electron
in a stationary potential. Because  the energy is conserved during such a
motion, the function
$K^{\cal D}$ can be factorized to the form
\begin{eqnarray} &&\displaystyle{K_{\epsilon}^{\cal D}\left(\omega;\ {\bf
p}_1, {\bf R}_1;
\ {\bf p}_2, {\bf R}_2\right) =  {\cal D}_{\epsilon}\left(\omega;\ {\bf
n}_1, {\bf R}_1;
\ {\bf n}_2, {\bf R}_2\right)\times} \nonumber\\
&&\displaystyle{\quad\quad\frac{2\pi}{\nu}\delta\left[\epsilon-{\cal
H}({\bf p}_1, {\bf R}_1)\right] \delta\left[\epsilon-{\cal H}({\bf p}_2,
{\bf R}_2)\right]}, 
\label{diffuson1}
\end{eqnarray} where diffuson ${\cal D}$ is a smooth  function of the
electron energy,
${\bf n}$ is the unit vector  along the momentum direction, ${\bf p} =
p_F{\bf n} ={\bf n}\sqrt{2m\left[\epsilon-U({\bf r})\right]}$
 and
$\nu = m/2\pi$ is the density of states.  Diffuson ${\cal D}_{\epsilon}$
is the solution of the equation
\begin{equation}
\left[-i\omega + \hat{L}_1\right]{\cal D}
 =\delta_{12}, \quad \delta_{12}\equiv 2\pi\delta({\bf n}_1-{\bf n}_2)
\delta({\bf R}_1-{\bf R}_2).
\label{KD1}
\end{equation}   It is important to emphasize that the diffuson ${\cal
D}$ is a solution of the Liouville equation and not of the diffusion
equation. In this sense, a more correct term for ${\cal D}$ is
``Liouvillon'',  however, we follow  the terminology accepted in the
theory of quantum disorder.

Let us consider the classical chaotic motion such that  the time of the
randomization of momentum direction is finite. At small $\omega$, which
corresponds to the averaging over
 time scale much larger than the time of the momentum randomization,
${\cal D}_{\epsilon}$  averaged over small region of its initial
conditions, satisfies the diffusion equation
\begin{equation} {\cal D}= 
\frac{1}{-i\omega - D\nabla^2},
\label{diffaverage}
\end{equation} where $D$ is the diffusion constant. The explicit relation
of $D$ to the characteristics of the potential $U$ can be found in the
limit of dilute scatterers
$l_{tr} \gg a$: in this limit the diffusion constant is given by
$D=v_F^2\tau_{tr}/2$. It is worth emphasizing that Eq.~(\ref{diffaverage})
itself  does not require such a small parameter, and it is always valid at
large spatial scales and small frequencies . We will ignore the possible
islands in the phase space  isolated from the rest of the system.

The semiclassical equation for function $K_{\epsilon}^{\cal C}$ from
Eq.~(\ref{mixedrepC}) is found in a similar fashion: in the absence of the
magnetic field and spin-orbit scattering it reads
\begin{eqnarray}
\left[-i\omega + \hat{L}_1\right]K_{\epsilon}^{\cal C}\left(\omega;\ {\bf
p}_1, {\bf R}_1;
\ {\bf p}_2, {\bf R}_2\right) = \label{KC}\\ 
2\pi\delta\left[\epsilon-{\cal H}({\bf p}_1, {\bf R}_1)\right]
(2\pi)^2\delta({\bf p}_1-{\bf p}_2)
\delta({\bf R}_1-{\bf R}_2).
\nonumber
\end{eqnarray} Function $K_{\epsilon}^{\cal C}$ can be  factorized as
\begin{eqnarray} &&\displaystyle{K_{\epsilon}^{\cal C}\left(\omega;\ {\bf
p}_1, {\bf R}_1;
\ {\bf p}_2, {\bf R}_2\right) =  {\cal C}\left(\omega;\ {\bf n}_1, {\bf
R}_1;
\ {\bf n}_2, {\bf R}_2\right)\times} \nonumber\\
&&\displaystyle{\quad\quad\frac{2\pi}{\nu}\delta\left[\epsilon-{\cal
H}({\bf p}_1, {\bf R}_1)\right] \delta\left[\epsilon-{\cal H}({\bf p}_2,
{\bf R}_2)\right]}. 
\label{Cooperon1}
\end{eqnarray} Here  Cooperon ${\cal C}_{\epsilon}$ is a smooth  function
of the  electron energy satisfying the equation
\begin{equation}
\left[-i\omega + \hat{L}_1\right]{\cal C}
 =
\delta_{12}.
\label{KC1}
\end{equation}

Similar to the diffuson, the Cooperon, averaged over small region of its
initial conditions, is a self-averaging quantity  at large distances and
small frequencies and in the absence of magnetic field and spin-orbit
scattering, it can be described by  the expression analogous to
Eq.~(\ref{diffaverage}):
\begin{equation} {\cal C}= 
\frac{1}{-i\omega - D\nabla^2}.
\label{cooperaverage}
\end{equation}

\subsection{Quantum corrections to classical probabilities.}
\label{sec:3.2}

So far, we considered the lowest classical approximation, in which the
classical probabilities were determined by the deterministic equations of
the first order. However, the potential $U$ contains not only the
classical smooth part which is taken into account by the Liouville
equation, but also the part, responsible for the small angle diffraction.
Quantum weak localization correction originates from the interference of
the diffracted electron waves. The interference of the waves diffracted
at different locations is added. It results, as we will show below, the
quantum correction ceases to depend upon the details of the diffraction
mechanism and becomes universal. The  only quantity which depends on the
diffraction angle is the time it takes to establish this universality. We
will show, see also Sec.~\ref{sec:2}, that the dependence of this time on
the diffraction angle is only logarithmical. Therefore, with the
logarithmic accuracy, we can include the effect of this diffraction into
the classical Liouville equation by any convenient method, provided that
we do it consistently for all the quantities and preserve the
conservation of the number of particles.

We will model the diffraction by adding the small amount of the quantum
small angle scatterers to the LHS of the Schr\"{o}dinger equation
(\ref{G}). The effect of these scatterers will be twofold: 1) They will
smoothen the sharp classical probabilities; 2) They will induce
interaction between the diffuson and Cooperon modes, that results in the
weak localization correction. Finally, the strength and the density of
these scatterers will be adjusted so that the angle at which the classical
probability is smeared during the travel to the distance $a$ is equal to
the genuine diffraction angle
$\sqrt{\lambda_F/a}$. This procedure is legitimate because, as we already
mentioned, the  dependence of weak localization correction on the
diffraction angle is only logarithmical.

It is worth emphasizing, that even though weak localization correction
takes its origin at the very short linear scale (ultraviolet cut-off),
the value of this correction at very large distances does not depend on
this cut-off at all. Such phenomena are quite typical in physics, ({\em
e.g.} in the theory of turbulence, theory of strong interaction, or in
the Kondo effect).

Let us now implement the procedure. Consider a single impurity located at
the point
${\bf s}$, and creating the potential $V({\bf  r})=V_0({\bf s - r})$, so
that the potential part of the Hamiltonian (\ref{Hamiltonian}) is now
given by
$U({\bf r}) + V({\bf r})$. The characteristic size of this potential,
$d$, is much larger than
$\lambda_F$ but much smaller than
$a$. Our goal is to find the correction to  Eqs.~(\ref{KD1})  and
(\ref{KC1}) in the second order of perturbation theory in potential $V$.
(Correction of the first order vanishes if ${\cal D}$ and ${\cal C}$ are
functions smooth on the spatial scale  $d$.) In this order, correction to
function (\ref{K}) has the form 
}
\wide{m}{ 
\begin{eqnarray} &&\displaystyle{\delta K\left( {\bf r}_1, {\bf r}_2,{\bf
r}_3,{\bf r}_4 \right) = \int d {\bf r}_5{\bf r}_6
\left[G^R({\bf r}_1,{\bf r}_5)V({\bf r}_5) G^R({\bf r}_5,{\bf r}_2)
G^A({\bf r}_3,{\bf r}_6)V({\bf r}_6) G^A({\bf r}_6,{\bf r}_4)
\right.+} \label{correctionK}\\ &&\displaystyle{\left.\quad  G^R({\bf
r}_1,{\bf r}_5)V({\bf r}_5) G^R({\bf r}_5,{\bf r}_6)V({\bf r}_6) G^R({\bf
r}_6,{\bf r}_2) G^A({\bf r}_3,{\bf r}_4)+ G^R({\bf r}_1,{\bf
r}_2)G^A({\bf r}_3,{\bf r}_5)V({\bf r}_5) G^A({\bf r}_5,{\bf r}_6) V({\bf
r}_6) G^R({\bf r}_6,{\bf r}_2)
\right],}
\nonumber 
\end{eqnarray} } 
where Green functions   are the solutions of
Eq.~(\ref{G}) without the impurity potential $V$. We will omit the energy
arguments in the Green  function, implying everywhere that the energies
for the retarded and advanced Green functions are
$\epsilon + \omega/2$ and $\epsilon - \omega/2$ respectively.

In order to find the correction to the diffuson, we consider  the points
$r_1, r_4$ and $r_2, r_3$ in Eq.~(\ref{correctionK}) which are close to
each other pairwise, perform the Fourier transform defined by
Eq.~(\ref{mixedrepD}), and express the RHS of Eq.~(\ref{correctionK}) in
terms of the diffusons and Cooperons. We demonstrate the calculation by
evaluating the second term in the RHS of Eq.~(\ref{correctionK}), let us
denote it by 
$\delta K_2$.

Consider the product $G^R({\bf r}_1,{\bf r}_5)G^A({\bf r}_3,{\bf r}_4)$,
points
$r_1, r_4$ are close to each other, but points $r_5, r_3$ are not. It
means that for calculation of such a product we can not use the
semiclassical approximation (\ref{semGreen}) for the RHS of
Eq.~(\ref{eqK}) but sill can use the expansion (\ref{Lioville})  for the
LHS of Eq.~(\ref{eqK}). Solving Eq.~(\ref{eqK}) with the help of
Eq.~(\ref{KD1}), we obtain
\wide{m}{
\begin{eqnarray} &\displaystyle{G^R({\bf r}_1,{\bf r}_5) G^A({\bf
r}_3,{\bf r}_4) = 
\frac{i}{\nu}\int d {\bf r}_7 {\bf r}_8
\int \frac{d {\bf p}_1}{(2\pi)^2}\frac{d {\bf p}_2}{(2\pi)^2} e^{i{\bf
p}_1({\bf r}_1- {\bf r}_4) - i{\bf p}_2({\bf r}_7- {\bf r}_8)} {\cal
D}\left({\bf n}_1, \frac{{\bf r}_1+ {\bf r}_4}{2};  {\bf n}_2, \frac{{\bf
r}_7+ {\bf r}_8}{2}\right)
\times }& \label{profGf}\\ &\displaystyle{
\left[G^R({\bf r}_7,{\bf r}_5)\delta({\bf r}_8-{\bf r}_3)- G^A({\bf r}_3,
{\bf r}_8)
\delta({\bf r}_7-{\bf r}_5)
\right]\delta\left[{\cal H}\left({\bf p}_1, \frac{{\bf r}_1+ {\bf r}_4}
{2}\right)  -{\cal H}\left({\bf p}_2, \frac{{\bf r}_7+ {\bf
r}_8}{2}\right)
\right] ,}&
\nonumber 
\end{eqnarray} } with ${\cal H}({\bf p}, {\bf r})$ being the Hamilton
function (\ref{HamiltonFunction}). We will omit the frequency argument 
in the diffusons and Cooperons, implying everywhere that it equals to
$\omega$.

We substitute Eq.~(\ref{profGf}) into the second term in the LHS of
Eq.~(\ref{correctionK}). We neglect the product of three retarded Green
functions because this product is a strongly oscillating function of its
arguments and vanishes after the averaging on a spatial scale larger than
$\lambda_F$. The remaining product $G^R({\bf r}_6,{\bf r}_2)G^A({\bf
r}_3,{\bf r}_8)$ is approximated by the expression similar to
Eq.~(\ref{profGf}) because points
${\bf r}_2$ and ${\bf r}_3$ are close to each other. Neglecting, once
again,  the product of two retarded Green functions and performing the
Fourier transform over the differences $r_1-r_4$ and $r_2-r_3$, we find

\wide{m}{ 
\begin{eqnarray} 
&&\displaystyle{\delta {\cal D}_2(1,2) =  -\int\!\! d
{\bf r}_3d {\bf r}_4d {\bf R}_3d {\bf R}_4
\int \!\!\frac{d {\bf p}_3}{(2\pi)^2}\frac{d {\bf p}_4}{(2\pi)^2}
e^{-i{\bf p}_3{\bf r}_3 + i{\bf p}_4{\bf r}_4} {\cal D}\left(1; 3\right)
{\cal D}\left(4; 2\right) G^R\left({\bf r}_3^+, {\bf r}_4^+) G^A({\bf
r}_4^-,{\bf r}_3^-\right)
\frac{V({\bf r}_3^+)V({\bf r}_4^+) } {2\pi\nu},}
\nonumber \\ 
&&\displaystyle{
\delta K_2\left( {\bf r}_1^+, {\bf r}_2^+, {\bf r}_2^-, {\bf
r}_1^-\right)= {2\pi\nu}
\int \frac{d {\bf n}_1}{(2\pi)}\frac{d {\bf n}_2}{(2\pi)} e^{ip_F{\bf
n}_1{\bf r}_1 - ip_F{\bf n}_2{\bf r}_2}
\delta {\cal D}_2\left(1,2\right) ,}
\label{almost}
\end{eqnarray}  Here we introduced the short hand 
notation $j\equiv ({\bf
n}_j, {\bf R}_j)$ and
${\bf r}_j^{\pm} ={\bf R}_j\pm \frac{{\bf r}_j}{2}$.

What remains is to find the semiclassical expression for the product
$G^RG^A$ in Eq.~(\ref{almost}). We notice that the points ${\bf r}_3^+,
{\bf r}_4^+$ lie within the radius of the potential $V(r)$. In order for
the product
$G^RG^A$  in Eq.~(\ref{almost}) not to vanish, the points
${\bf r}_3^-, {\bf r}_4^-$ must be close to the points ${\bf r}_3^+, {\bf
r}_4^+$. Because all the four points are close to each other, one can
write, cf. Eq.~(\ref{mixedrepC}),
\begin{eqnarray} G^R\left({\bf r}_3^+, {\bf r}_4^+) G^A({\bf r}_4^-,{\bf
r}_3^-\right) = \nu^2
\int\! {d {\bf n}_4}{d {\bf n}_5}\ \theta\ [{\bf n}_4({\bf r}_3^+  -{\bf
r}_4^+)]\
 \theta\ [{\bf n}_5({\bf r}_4^- -{\bf r}_3^-)]\ e^{ip_F{\bf n}_4({\bf
r}_3^+ -{\bf r}_4^+)+ ip_F{\bf n}_5({\bf r}_3^--{\bf r}_4^-)} + &&
\nonumber\\
\frac{\nu}{2\pi}
\int {d {\bf n}_4}{d {\bf n}_5} e^{ip_F{\bf n}_4({\bf r}_3^+ -{\bf r}_4^-
)+ ip_F{\bf n}_5({\bf r}_3^--{\bf r}_4^+)}{\cal C}
\left({\bf n}_4,\frac{{\bf r}_3^+ +{\bf r}_4^-}{2};{\bf n}_5,
\frac{{\bf r}_3^--{\bf r}_4^+}{2}\right).&&
\label{mprod}
\end{eqnarray} }

Here, the first term is the explicitly separated contribution of the short
straight line trajectories connecting points $r_1, r_2$ and $r_3, r_4$. 
These short trajectories can be well described by the Cooperon or by the
diffuson.  The second term describes the contribution of all the other
trajectories connecting these points. It can be shown by explicit
calculation that the representation of Eq.~(\ref{mprod}) in terms of the
diffuson only would lead to the loss of this second term. This is because
the Cooperon describes the interference effects corresponding to the
oscillating part of the diffuson which is lost in the semiclassical
approximation (\ref{KD1}).

Now, we are ready to find the correction coming from the single quantum
scatterer. We substitute Eq.~(\ref{mprod}) into Eq.~(\ref{almost}) and
perform the integration while neglecting the dependence of the diffusons
and Cooperon on their spatial coordinates on the scale of the order of the
scatterer size. We consider the remaining two terms in
Eq.~(\ref{correctionK}) in a similar manner. The  overall result is
\begin{eqnarray} &&\displaystyle{\delta {\cal D} = \delta {\cal D}^{St} +
\delta{\cal D}^{I},}
\label{singleimpurity}\\ &&\displaystyle{\delta {\cal D}^{St}(1,2) = \int
d3d4 {\cal P}_s(3,4) {\cal D}(1,3)\left[{\cal D}(4,2)-{\cal
D}(3,2)\right],}\nonumber \\ &&\displaystyle{\delta {\cal D}^{I}(1,2) =
\int d3d4 {\cal P}_s(3,4)
\frac{{\cal C}(3,\bar{4})}{2\pi\nu}\times} \nonumber \\
&&\displaystyle{\hspace*{1.5cm}
\left[{\cal D}(1,3)-{\cal D}(1,4)\right]
\left[{\cal D}(\bar{3},2)-{\cal D}(\bar{4},2)\right].}
\nonumber
\end{eqnarray} Here we use the short hand notation $j\equiv ({\bf n}_j,
{\bf R}_j)$,  integration over the  phase space on the energy shell is
defined as $dj
\equiv d{\bf n}_jd{\bf R}_j/2\pi$, the time reversed coordinate $\bar{j}$
is given by
$\bar{j}\equiv (-{\bf n}_j, {\bf R}_j)$, and the kernel ${\cal P}$
describing the scattering by an impurity is 
\[ {\cal P}_s(1,2)\! =\! 2\pi\nu\delta({\bf s}- {\bf R}_1)
\delta({\bf s}- {\bf R}_2)
\left|\int\! d{\bf r}e^{ip_F{\bf r}({\bf n}_1-{\bf n}_2)} V({\bf r})
\right|^2\!.
\] 

The first term $\delta D^{St}$ in Eq.~(\ref{singleimpurity}) coincides
with that obtained for otherwise free moving electrons. The second term,
$\delta D^I$, describes the interference effect arising because the
chaotically moving in classical potential
$U(r)$ electron may return to the vicinity of the impurity one more time.

The  correction to the Cooperon due to the single impurity can be obtained
from Eq.~(\ref{correctionK}) by considering close pairs $r_1, r_3$ and
$r_2, r_4$; this results in the expression similar to
Eq.~(\ref{singleimpurity}) with the replacement
${\cal D} \leftrightarrow {\cal C}$.

So far, we considered the correction due to a single weak impurity. If the
number of these impurities is large, we can, in the lowest approximation,
consider the contributions from the different impurities independently of
each other, by the substitution to the RHS of Eqs.~(\ref{singleimpurity})
of the diffusons and Cooperons renormalized by all the other impurities.
As the result, we arrive to the Boltzmann-like equations for the diffuson
and Cooperon

\wide{m}{ 
\begin{mathletters}
\label{Boltzmann}
\begin{eqnarray} &&\displaystyle{\left[-i\omega + \hat{L}_1\right]{\cal
D}(1,2)=
 \delta_{12} + \sum_{\bf s}\int d3{\cal P}_s(1,3)
\left\{\left[{\cal D}(3,2)-{\cal D}(1,2)\right] + 
\left[{\cal D}(\bar{1},2)-{\cal D}(\bar{3},2)\right] \frac{{\cal
C}(1,\bar{3})+{\cal C}(3,\bar{1})}{2\pi\nu}
 \right\}}
\label{BoltzmannD}\\  &&\displaystyle{\left[-i\omega +
\hat{L}_1\right]{\cal C}(1,2)=
 \delta_{12} + \sum_{\bf s}\int d3{\cal P}_s(1,3)
\left\{\left[{\cal C}(3,2)-{\cal C}(1,2)\right] + 
\left[{\cal C}(\bar{1},2)-{\cal C}(\bar{3},2)\right] \frac{{\cal
D}(1,\bar{3})+{\cal D}(3,\bar{1})}{2\pi\nu} \right\}}
\label{BoltzmannC}
\end{eqnarray}
\end{mathletters} } where the notation for the coordinates $j, \bar{j}$
was introduced after Eq.~(\ref{singleimpurity}), and the $\delta$ -
symbol was defined in Eq.~(\ref{KD1}).

\begin{mathletters}
\label{diffusion} Assuming that the distribution of the quantum
impurities is uniform with the density
$n_i$, we can make the continuous approximation and replace $\sum_{\bf s}
\to n_i\int d{\bf s}$ in the RHS of Eqs.~(\ref{Boltzmann}). Finally,
taking into account that the scattering angle is small, we reduce
Eqs.~(\ref{Boltzmann}) to a differential form. Equation
(\ref{BoltzmannD}) becomes 
\begin{equation}
\left[-i\omega +
\hat{L}_1-\frac{1}{\tau_q}\frac{\partial^2}{\partial\phi^2_1}\right]{\cal
D} (1) =\delta_{12}-\frac{\partial}{\partial\phi_1}
\frac{{\cal C}(1,\bar{1})}{\pi\nu\tau_q }
\frac{\partial}{\partial\phi_1}{\cal D}(\bar{1})
\label{diffusionD}
\end{equation} Here, angle $\phi_j$ is defined so that
${\bf n}_j = (\cos \phi_j,\sin\phi_j)$, the notation for the coordinates 
$j\equiv ({\bf n}_j, {\bf R}_j),\ \bar{j}\equiv (-{\bf n}_j, {\bf R}_j)$,
is the same as in Eq.~(\ref{singleimpurity}), and the $\delta$ - symbol
was defined in Eq.~(\ref{KD1}). The second argument is the same for all
the diffusons in  Eq.~(\ref{diffusionD}) and that is why we omitted it.
Analogously, Eq.~(\ref{BoltzmannC}) reduces to
\begin{equation}
\left[-i\omega +
\hat{L}_1-\frac{1}{\tau_q}\frac{\partial^2}{\partial\phi^2_1}\right]
{\cal C}(1) =\delta_{12}-\frac{\partial}{\partial\phi_1}
\frac{{\cal D}(1,\bar{1})}{\pi\nu\tau_q }
\frac{\partial}{\partial\phi_1}{\cal C}(\bar{1})
\label{diffusionC}
\end{equation} The second argument is the same for all the Cooperons in 
Eq.~(\ref{diffusionC}) and it is omitted.
\end{mathletters} Quantum transport life time in Eqs.~(\ref{diffusion})
is given by
\[
\frac{1}{\tau_q} =\! 2\pi\nu n_i \int \frac{d\phi}{2\pi}\frac{\phi^2}{2}
\left|\int\! d{\bf r}e^{ip_F{\bf n}{\bf r}\phi} V({\bf r}) \right|^2\!.
\] 

Equations (\ref{diffusion}) describe how the classical Liouville equation
changes under the effect of the small angle scattering (diffraction). We
see that the quantum effects result in  two contributions to the Liouville
equation. First contribution provides the angular diffusion and, thus, it
leads to the smearing of the sharp classical probabilities. Usually, for
the calculation of the transport coefficients, such as the diffusion
constant or the conductivity, the averaging over the initial and final
coordinates is performed anyway. Therefore, the angular diffusion itself
provides only negligible correction to the classical transport
coefficients which are controlled by classical potential $U$. On the
contrary, the second contribution giving the quantum correction [last
terms in the RHS of Eq.~(\ref{diffusionD})] is proportional to the
classical probability
$C(\bar{1},1)$ where the initial and finite points of the phase space are
related by the time inversion. In the absence of the spreading due to the
angular diffusion, $\tau_q \to \infty$, this probability vanishes
identically, see Sec.~\ref{sec:2}. In order to obtain the correction at
finite time (or finite frequency), one must keep $\tau_q$ finite even in
the final results.

Let us estimate the value one should ascribe to $\tau_q$ for the
description of the diffraction effects in the system.  As we already
discussed, for the calculation with the logarithmic accuracy, we do not
need the numerical coefficient. The parametric dependence of $\tau_q$ can
be established by using the following argument. Consider two independent
electrons, starting with the same initial conditions. If there were no
diffraction, they would propagate together forever. Due to the angular
diffusion (diffraction), the directions of these trajectories deviates
first $\propto \sqrt{t}$ and then exponentially,
$
\frac{d\langle \delta\phi^2\rangle}{dt}\approx 2\lambda\langle
 \delta\phi^2\rangle +
\frac{1}{2\tau_q}
$,  where angle $\delta\phi$ stands for the angle between the momenta of
two electrons, and $\lambda$ is the Lyapunov exponent. It yields $\langle
\delta\phi^2(t)\rangle \approx (4\lambda\tau_q)^{-1}\left( e^{2\lambda
t}-1\right)$. Thus, the characteristic time during which the angular
diffusion switches to the exponential growth is always $t_e \simeq
1/\lambda$. On the other hand, quantum spreading of the wave packet during
this time interval is given by
$\delta x^2 \simeq \lambda_F v_F t_e$. Taking into account the relation
$\delta x\simeq \phi t_e v_F$, we find $t_e/\tau_q \simeq
\lambda_F/(v_Ft_e)$. It yields the estimate for the quantum transport time
entering into Eqs.~(\ref{diffusion}) corresponding to the small angle
diffraction
\begin{equation}
\frac{1}{\tau_q}\simeq \lambda^2\frac{\lambda_F}{v_F}.
\label{tauestem}
\end{equation}

It is important to emphasize that the very same $\tau_q$ enters into the
angular diffusion term and into the diffuson-Cooperon interaction. This
circumstance is extremely crucial for the universality of the quantum
correction at large time ($\omega \to 0$), even though parameter $\tau_q$
itself does not enter into the result, see Sec.~\ref{sec:4}.

\begin{mathletters}
\label{Mainresult} Let us now turn to the calculation of the lowest
quantum correction to the diffuson. Taking into account the last term in
the RHS of Eq.~(\ref{diffusionD}) in the first order of perturbation
theory, we obtain
\begin{eqnarray} &&{\cal D}(1,2) = {\cal D}^0(1,2)+\Delta{\cal D}(1,2);
\label{mainresultg}\\ &&\Delta{\cal D}(1,2)= \int d3 \frac{{\cal
C}^0(3,\bar{3})} {\pi\nu\tau_q}
\frac{\partial {\cal D}^0(1,3)}{\partial\phi_3}
\frac{\partial {\cal D}^0(\bar{3},2)}{\partial\phi_3};
\label{mainresult0}\\ &&\left[-i\omega +
\hat{L}_1-\frac{1}{\tau_q}\frac{\partial^2}
{\partial\phi^2_1}\right]{\cal D}^0(1,2) =\delta_{12};
\label{mainresultD}\\ &&\left[-i\omega +
\hat{L}_1-\frac{1}{\tau_q}\frac{\partial^2}{\partial\phi^2_1}\right]
{\cal C}^0(1,2) =\delta_{12},
\label{mainresultC}
\end{eqnarray} where $j\equiv ({\bf n}_j, {\bf R}_j)$,  integration over
the  phase space on the energy shell is defined as 
$dj \equiv d{\bf n}_jd{\bf R}_j/2\pi$, the time reversed coordinate
$\bar{j}$ is given by
$\bar{j}\equiv (-{\bf n}_j, {\bf R}_j)$, and the $\delta$ - symbol was
 defined in Eq.~(\ref{KD1}).
\end{mathletters}

Equation~(\ref{mainresult0}) can be rewritten in a different form. Even
though more lengthy than Eq.~(\ref{mainresult0}), this form  turns out to
be more convenient for further applications:  
\begin{eqnarray} &&\Delta{\cal D}(1,2) =  {\cal D}^0(1,\bar{2})
\frac{{\cal C}^0(\bar{2},2)}{2\pi\nu} +
\frac{{\cal C}^0(1,\bar{1})}{2\pi\nu}{\cal D}^0(\bar{1},2) 
\label{mainresult}\\ &&+\int d3 
 {\cal D}^0(1,3){\cal D}^0(\bar{3},2)
\left[ 2i\omega-
\hat{L}_3 +\frac{1}{\tau_q}\frac{\partial^2}{\partial\phi^2_3}
\right]
\frac{{\cal C}^0(3,\bar{3})}{2\pi\nu}.
\nonumber
\end{eqnarray} In order to derive Eq.~(\ref{mainresult}) from
Eq.~(\ref{mainresult0}),
 we subtracted from the RHS of Eq.~(\ref{mainresult}) the expression
\[
\int d3\hat{L}_3\left[\frac{{\cal C}(3,\bar{3})}{2\pi\nu}{\cal
D}(1,3){\cal D}(\bar{3},2)\right],
\] which vanishes because integrand is the total derivative along the
classical trajectory. Then, we integrated Eq.~(\ref{mainresult0}) by
parts and,
 with the help of Eq.~(\ref{mainresultD}),  we arrived to
Eq.~(\ref{mainresult}).

Equations (\ref{mainresult0}) and (\ref{mainresult}) are the main results
of this section. They give the value of the lowest quantum correction to
the classical correlator in terms of the non-averaged solutions of the 
Liouville equation (with small angular diffraction added) for a given
system.  Besides the found correction, there exist the other corrections
[{\em e.g.} from the higher terms in the expansion (\ref{Lioville})],
however, Eqs.~(\ref{mainresult0}) and (\ref{mainresult}) are dominant at
low frequencies. The quantum mesoscopic fluctuations are  neglected in
Eqs.~(\ref{mainresult0}) and (\ref{mainresult}), which implies either the
temperature is high enough or the averaging over the position of the
Fermi level is performed. Then, if the relevant time and spatial scales
are large, the quantum correction becomes a self-averaging quantity
expression for which will be obtained in the next section.  

\section{Averaged quantum corrections}
\label{sec:4} We will consider the quantum correction at large distance
and time scales. In this case, the classical probability does not depend
on the direction of the momentum and it is given by
Eq.~(\ref{diffaverage}). Our goal now is to find the expression for the
quantum correction in the same approximation. We will bear in mind the
systems in which the diffusion constant is large enough,
$D/av_F \gtrsim 1$.  It is the case for the antidot arrays. The
conductance of the net of the ballistic cavities requires a separate
consideration.

For the calculation  we  use Eq.~(\ref{mainresult}).  While performing
the averaging, we make use of the fact that the Cooperon part of the
expression can be averaged independently on the diffuson part. This is
because the classical trajectories corresponding to these quantities lie
essentially in the different spatial regions, (see e.g. Fig.~\ref{fig:2}b,
where segments $iB$ and $fB$ correspond to diffusons and segment
$BEB$ corresponds to the Cooperon) and, therefore, they are governed by
the different potentials and are not correlated. Performing such an
averaging, we obtain  from Eq.~(\ref{mainresult}):
\begin{eqnarray} &&\Delta{\cal D}(1,2) = 
\left[\raisebox{0mm}[3mm][3mm]{$
\langle {\cal D}^0(1,\bar{2}) \rangle +
\langle {\cal D}^0(\bar{1},{2}) \rangle
 +$}
\right.\\
\label{mainresultav} &&\hspace*{2cm}
\left. 2i\omega \int d3 \langle{\cal D}^0(1,3){\cal
D}^0(\bar{3},2)\rangle\right]
\displaystyle{\frac{\langle{\cal
C}^0(1,\bar{1})\rangle}{2\pi\nu}.}\nonumber
\end{eqnarray} where $\langle\dots\rangle$ stands  for the averaging
either over the realization of potential $U$ or over the position of the
``center of mass'' of the Cooperon and diffuson. The last two terms in
brackets in Eq.~(\ref{mainresult}) vanish after averaging because the
averaged cooperon does not depend on the coordinates
${\bf n}_3, {\bf R}_3$.

On the other hand, as we have already explained in Sec.~\ref{sec:2}, the
correlations in the motion of both ends of the Cooperon can not be
neglected. The same is also true about the correlation between motion of
the ends
$3$ and $\bar{3}$ in  the third term of Eq.~(\ref{mainresult}). In what
follows, we will separate the description of the problem into the Lyapunov
and diffusion regions. It will be done in subsections \ref{sec:4.2} and
\ref{sec:4.2d} for the Cooperon and diffusons respectively, and the
resulting correction to the conductivity will be found in subsection
\ref{sec:4.1}. The description of the Lyapunov region is presented in
subsections
\ref{sec:4.3} and \ref{sec:4.4}.

\subsection{Cooperon in the diffusive and Lyapunov regions} 
\label{sec:4.2} 

 In order to find $\langle{\cal C}^0(1,\bar{1})\rangle$ we consider more
general quantity ${C}(\phi,\rho)$ defined as
\begin{eqnarray} &&{C}(\phi,\rho) = 
\frac{1}{S}\int \frac{d{\bf R}d{\bf n}}{2\pi}{\cal C}^0
\left( {\bf n}^+, {\bf R}^-; -{\bf n}^-, {\bf R}^+ \right),
\label{Crhophi}\\ && {\bf n}^\pm = {\bf n} \cos\frac{\phi}{2} \pm
\left[{\bf n}\times{\bf l}_z\right]\sin\frac{\phi}{2}, \ {\bf R}^\pm =
{\bf R} \pm  
 \frac{\rho}{2}\left[{\bf n}^\pm\times{\bf l}_z\right], \nonumber 
\end{eqnarray} where $S$ is the area of the sample, and ${\bf l}_z$ is
the unit vector perpendicular to the plane. Function 
${C}(0,0)$ coincides with the necessary quantity $\langle{\cal
C}^0(1,\bar{1})\rangle$.

It is easy to find ${C}(\phi,\rho )$ in the diffusion region.  At $\rho
\gtrsim l_{tr}$, it  is given  by
\begin{equation} {C}\left(\phi,
\rho\right) = \frac{1}{-i\omega - D\nabla^2_\rho}. 
\label{diffregion} 
\end{equation}
 At $\rho < \sqrt{D/\omega}$ the Cooperon depends only logarithmically on
$\rho$ and at $a\lesssim\rho
\lesssim l_{tr}$, it becomes independent of $\rho$. With the logarithmic
accuracy, we have
\begin{equation} { C}\left(\phi,
\rho\right) \approx \frac{1}{4\pi D} \ln
\left(\frac{1}{\omega\tau_{tr}}\right), \quad \rho \gtrsim a.
\label{diffregionb} 
\end{equation} Equation (\ref{diffregionb}) serves as the boundary
condition for 
${C}\left(\phi,\rho\right)$ at the boundary between the diffusive and
Lyapunov regions:
\begin{equation} {C}\left(\phi,
\rho = a\ {\rm sign}\phi\right) \approx \frac{1}{4\pi D} \ln
\left(\frac{1}{\omega\tau_{tr}}\right).
\label{boundarycondition} 
\end{equation}  Meaning of Eq.~(\ref{boundarycondition}) is that both
ends of the Cooperon enter into the Lyapunov region with the random
momenta, and thus the probability of this entrance is given by the
solution of the diffusion equation.

The next step is to find ${C}(\phi,\rho)$ in the Lyapunov region. We add 
to  Eq.~(\ref{mainresultC}), the equation conjugate to it, which gives
\begin{equation}
\left[-2i\omega + \hat{L}_1 + \hat{L}_2 -
\frac{1}{\tau_q}\frac{\partial^2}{\partial
\phi_1^2}-\frac{1}{\tau_q}\frac{\partial^2}{\partial \phi_2^2}
\right]{\cal C}^0(1,\bar{2})
 = 2\delta_{1\bar{2}}.
\label{KC2}
\end{equation} Formula~(\ref{KC2}) enables us to find the equation for
quantity ${\cal C}^0
\left({\bf n},{\bf R};\phi,\rho\right)\equiv {\cal C}^0
\left( {\bf n}^+, {\bf R}^-; -{\bf n}^-, {\bf R}^+ \right)$  from
Eq.~(\ref{Crhophi}). Expanding potential $U$ up to the first order in
$\rho$,  and using the fact that the angle 
$\phi$ is small, we obtain 
\begin{equation}
\left[-2i\omega + \hat{L}_c+\hat{L}_r -\frac{1}{2\tau_q}
\frac{\partial^2}{\partial\phi^2}\right]{\cal C}^0\left({\bf n}, {\bf R};
\phi,
\rho
\right) = 0.
\label{LyapunovRegion}
\end{equation} Here operator
\begin{equation}
\hat{L}_c = v_F{\bf n}\cdot
\frac{\partial}{\partial {\bf R}} - \frac{\partial U({\bf R})}{\partial
{\bf R}}
\cdot
 \frac{\partial}
               {\partial {\bf P}}
\label{L+approx}
\end{equation} describes the motion of the ``center of mass'' of the
Cooperon along a classical trajectory and operator $\hat{L}_r$
characterizes how the distance between the ends changes  in a course of
this motion:
\begin{equation}
\hat{L}_r = -v_F\phi\frac{\partial}{\partial \rho} +
\frac{\partial^2U}{p_F\partial R_\perp^2}\rho\frac{\partial}{\partial
\phi}.
\label{L-approx}
\end{equation} with $ R_\perp$ being the projection of ${\bf R}$  onto
the direction perpendicular to ${\bf n}$. In Eq.~(\ref{LyapunovRegion}),
we neglected the effect of the angular diffusion on the motion of the
center of mass because the averaging over the position of the center of
mass ${\bf n, \ R}$ is performed in Eq.~(\ref{Crhophi}) anyway.

Now, we have to find function  $C(\rho,\phi)$ in the Lyapunov region,  
satisfying the boundary condition given by Eq.~(\ref{boundarycondition})
and consistent with Eqs.~(\ref{Crhophi}) and (\ref{LyapunovRegion}).
Solution can be represented in a compact form analogous to
Eq.~(\ref{integral})
\begin{equation} {C}\left(\phi,
\rho\right) = \frac{w\left(\omega; \phi, \rho \right)}{4\pi D} \ln
\left(\frac{1}{\omega\tau_{tr}}\right).
\label{solution1}
\end{equation} Function $w\left(\omega; \phi, \rho \right)$ is defined as
\begin{equation} w\left(\omega; \phi, \rho \right) = 
\frac{1}{S}\int \frac{d{\bf R}d{\bf n}} {2\pi}W\left(\omega; {\bf n},
{\bf R}; \phi,
\rho
\right),
\label{wbar}
\end{equation} where $S$ is the area of the sample and $W$ is the 
solution of the equation
\begin{equation}
\left[-2i\omega + \hat{L}_c+\hat{L}_r -\frac{1}{2\tau_q}
\frac{\partial^2}{\partial\phi^2}\right] W\left(\omega; {\bf n}, {\bf R};
\phi,
\rho
\right) = 0,
\label{wequation}
\end{equation} supplied with the boundary condition 
\begin{equation} W\left(\omega; {\bf n}, {\bf R}; \phi,
\rho = a\ {\rm sign}\phi\right)=1.
\label{WBC}
\end{equation} The necessary quantity $\langle{\cal
C}^0(1,\bar{1})\rangle$ is,  thus, found by putting $\rho,\phi=0$ in
Eq.~(\ref{solution1})
\begin{equation}
\langle{\cal C}^0(1,\bar{1})\rangle = 
\frac{w\left(\omega; 0, 0 \right)}{4\pi D}
\ln
\left(\frac{1}{\omega\tau_{tr}}\right).
\label{cfinal}
\end{equation}

\subsection{Diffusons in the diffusive and Lyapunov regions.} 
\label{sec:4.2d} In this subsection we find the average $\int
d3\langle{\cal D}^0(1,3){\cal D}^0(\bar{3},2)\rangle$ entering into
Eq.~(\ref{mainresultav}). We use the procedure  similar to the
calculation of the Cooperon in the previous subsection. We consider more
general quantities $M,\ {\cal M}$ defined as
\begin{eqnarray} &&{\cal M}(1,2; {\bf n, R}; \phi,\rho)={\cal D}^0
\!
\left(1; -{\bf n}^-, {\bf R}^+\ \right){\cal D}^0\!\left({\bf n}^+,  {\bf
R}^-; 2 \right)
\nonumber \\ &&\displaystyle{M(1,2; \phi,\rho)\! =\!\! 
\int\!\! \frac{d{\bf R}d{\bf n}}{2\pi}\langle {\cal M}(1,2; {\bf n, R};
\phi,\rho)\rangle,}
\label{Drhophi}
\end{eqnarray} where the coordinates ${\bf n}^\pm, {\bf R}^\pm$ are
defined in Eq.~(\ref{Crhophi}). Function 
$M(1,2; 0,0)$ coincides with the necessary quantity $\int d3\langle{\cal
D}^0(1,3){\cal D}^0(\bar{3},2)\rangle$.

In the diffusive region $\rho \gtrsim a$ two diffusons are governed by the
different potentials and, therefore, can be averaged independently; each
of them is given by Eq.~(\ref{diffaverage}). Furthermore,  if
$\rho
\ll \sqrt{D/\omega}$, function $M(1,2; \phi,\rho)$ becomes independent of
$\rho,\phi$ and it is given by 
\begin{equation} M(1,2; \phi,\rho) = \int d3\langle{\cal
D}^0(1,3)\rangle\langle{\cal D}^0(\bar{3},2)\rangle.
\label{ddif}
\end{equation}  Equation (\ref{ddif}) serves as the boundary condition 
for  $M(1,2; \phi,\rho)$ at the boundary between the diffusion and
Lyapunov regions:
\begin{equation}
\!\!M\left(1,2;\phi,
\rho = a\ {\rm sign}\phi\right)\! =\!\int\!\! d3\langle{\cal
D}^0(1,3)\rangle \langle{\cal D}^0(\bar{3},2)\rangle.
\label{Mbc} 
\end{equation}  Meaning of Eq.~(\ref{Mbc}) is that the ends of both
diffusons enter into the Lyapunov region with the random momenta.

The next step is to find ${M}(1,2;\phi,\rho)$ in the Lyapunov region. It
follows from  Eq.~(\ref{mainresultD}), that the product of two diffusons 
${\cal D}^0
\!
\left(1; \bar{3}\right){\cal D}^0\!\left( 4; 2
\right)$ satisfies the equation
\begin{eqnarray} &&\left[-2i\omega + \hat{L}_3 + \hat{L}_4 -
\frac{1}{\tau_q}\frac{\partial^2}{\partial
\phi_3^2}-\frac{1}{\tau_q}\frac{\partial^2}{\partial \phi_4^2}
\right]{\cal D}^0
\!
\left(1; \bar{3}\right){\cal D}^0\!\left( 4; 2
\right)
\nonumber\\ &&\hspace*{2cm}=\delta_{1\bar{3}}{\cal D}^0\!\left( 4;
2\right) +
\delta_{2\bar{4}}{\cal D}^0\!\left(1; \bar{3}\right).
\label{KD2}
\end{eqnarray}
 Equation~(\ref{KD2}) enables us to find the equation for  quantity
${\cal M}$ from Eq.~(\ref{Drhophi}). We expand the potential $U$ up to
the first order in $\rho$,  and use the fact that the angle 
$\phi$ is small. This yields
\begin{eqnarray} &&\left[-2i\omega + \hat{L}_c+\hat{L}_r
-\frac{1}{2\tau_q}
\frac{\partial^2}{\partial\phi^2}\right] {\cal M}
\left(1,2;{\bf n},{\bf R};\phi,\rho\right)=
\nonumber\\ &&\hspace*{0.2cm}2\pi\delta\left({\bf n}_1+{\bf n}^-\right)
\delta\left({\bf R}_1-{\bf R}^+\right) {\cal D}^0\!\left( {\bf n}^+, {\bf
R}^- ; 2\right) + \nonumber\\ &&\hspace*{0.2cm} 2\pi\delta\left({\bf
n}_2-{\bf n}^+\right)
\delta\left({\bf R}_2-{\bf R}^-\right) {\cal D}^0\!\left(1;- {\bf n}^-,
{\bf R}^+ \right),
\label{KD3}
\end{eqnarray} where the operators  $\hat{L}_c$ and $\hat{L}_r$ are
defined in Eqs.~(\ref{L+approx}) and (\ref{L-approx}) respectively. In
Eq.~(\ref{KD3}), we neglected the effect of the angular diffusion on the
motion of the center of mass because the averaging over the position of
the center of mass ${\bf n, \ R}$ is performed in Eq.~(\ref{Drhophi}).

We have to find function  $M(1,2;\rho,\phi)$ in the Lyapunov region,  
satisfying the boundary condition given by Eq.~(\ref{Mbc}) and consistent
with Eqs.~(\ref{Drhophi}) and (\ref{KD3}). We represent functions 
$M, {\cal M}$ as the sum of two terms $M=M_1 + M_2,\
 {\cal M}={\cal M}_1 + {\cal M}_2$,
\begin{equation} M_i(1,2; \phi,\rho)\! =\!\! 
\int\!\! \frac{d{\bf R}d{\bf n}}{2\pi}\langle {\cal M}_i(1,2; {\bf n, R};
\phi,\rho)\rangle,
\label{mm}
\end{equation} for $i=1,2$.
 Function ${\cal M}_1$ is a solution of the inhomogeneous equation
\begin{eqnarray} &&\left[-2i\omega + \hat{L}_c+\hat{L}_r
-\frac{1}{2\tau_q}
\frac{\partial^2}{\partial\phi^2}\right] {\cal M}_1
\left(1,2;{\bf n},{\bf R};\phi,\rho\right)=
\nonumber\\ &&\hspace*{0.2cm}2\pi\delta\left({\bf n}_1+{\bf n}^-\right)
\delta\left({\bf R}_1-{\bf R}^+\right) {\cal D}^0\!\left( {\bf n}^+, {\bf
R}^- ; 2\right) + \nonumber\\ &&\hspace*{0.2cm} 2\pi\delta\left({\bf
n}_2-{\bf n}^+\right)
\delta\left({\bf R}_2-{\bf R}^-\right) {\cal D}^0\!\left(1;- {\bf n}^-,
{\bf R}^+ \right),
\label{KD4}
\end{eqnarray} without any boundary conditions imposed and function 
${\cal M}_2$ is the solution of the homogeneous equation
\begin{equation}
\left[-2i\omega + \hat{L}_c+\hat{L}_r -\frac{1}{2\tau_q}
\frac{\partial^2}{\partial\phi^2}\right] {\cal M}_2
\left(1,2;{\bf n},{\bf R};\phi,\rho\right)=0,
\label{KD5}
\end{equation} with the boundary condition
\begin{eqnarray} &&M_2\left(1,2;\phi,
\rho = a\ {\rm sign}\phi\right) = \!\int\!\! d3\langle{\cal
D}^0(1,3)\rangle \langle{\cal D}^0(\bar{3},2)\rangle  - \nonumber \\
&&\hspace*{1.6cm} M_1\left(1,2;\phi,
\rho = a\ {\rm sign}\phi\right).
\label{Mbc2}
\end{eqnarray} First, we find function $M_1$. We integrate both sides of
Eq. (\ref{KD4})  over
${\bf R}, {\bf n}$ and average them. This gives
\begin{eqnarray} &&\left[-2i\omega  -\frac{1}{2\tau_q}
\frac{\partial^2}{\partial\phi^2}\right]  M_1(1,2;\rho, \phi)
+\label{KD6}\\ &&\int\!\! \frac{d{\bf n}d{\bf
R}}{2\pi}\langle\hat{L}_r{\cal M}_1
\left(1,2;{\bf n},{\bf R};\phi,\rho\right)\rangle\! =\!
\langle{\cal D}^0\!\left(\bar{1}; 2\right)\rangle\! + \!
\langle{\cal D}^0\!\left(1; \bar{2}\right)\rangle.
\nonumber
\end{eqnarray} Calculating the RHS of Eq.~(\ref{KD6}), we neglect $\rho
\lesssim a \ll
\sqrt{D\omega}$ in the arguments of the averaged diffusons.    Right hand
side of Eq.~(\ref{KD6}) is independent on $\rho$ and $\phi$. Therefore, we
can seek for the function $M_1(\rho, \phi)$ also independent of $\rho,
\phi$. The last term in the LHS of Eq.~(\ref{KD6}), then, vanishes and we
obtain
\begin{equation} M_1(1,2;\rho, \phi) = \frac{\langle{\cal
D}^0\!\left(\bar{1}; 2\right)
\rangle\!  + \!
\langle{\cal D}^0\!\left(1; \bar{2}\right)\rangle}{-2i\omega}.
\label{KD7}
\end{equation} Substituting Eq.~(\ref{KD7}) into Eq.~(\ref{Mbc2}), we
find the boundary condition for the function $M_2$
\begin{eqnarray} &&M_2\left(1,2;\phi,
\rho = a\ {\rm sign}\phi\right) = \!\int\!\! d3\langle{\cal
D}^0(1,3)\rangle \langle{\cal D}^0(\bar{3},2)\rangle  - \nonumber \\
&&\hspace*{1.6cm}
\frac{\langle{\cal D}^0\!\left(\bar{1}; 2\right)
\rangle\!  + \!
\langle{\cal D}^0\!\left(1; \bar{2}\right)\rangle}{-2i\omega}
\label{Mbc3}
\end{eqnarray} Equation (\ref{KD5}), supplied with the boundary condition
(\ref{Mbc3}), is similar to  Eqs.~(\ref{LyapunovRegion}) and
(\ref{boundarycondition}) for the Cooperon considered in the previous
subsection.   Thus, we use Eq.~(\ref{solution1}) to obtain 
\begin{eqnarray} &&M_2\left(1,2;\phi,
\rho \right) = w(\omega; \phi,\rho)\left[
\int\!\! d3\langle{\cal D}^0(1,3)\rangle \langle{\cal
D}^0(\bar{3},2)\rangle - \right.
\nonumber\\ &&\hspace*{2.6cm}\left.
\frac{\langle{\cal D}^0\!\left(\bar{1}; 2\right)
\rangle\!  + \!
\langle{\cal D}^0\!\left(1; \bar{2}\right)\rangle}{-2i\omega}
\right],
\label{M2solution}
\end{eqnarray} where function $w$ is defined by Eq.~(\ref{wbar}).

The necessary quantity $\int d3\langle{\cal D}^0(1,3){\cal
D}^0(\bar{3},2)\rangle$ is, thus, found by summing the contributions
(\ref{KD7}) and (\ref{M2solution}) and putting 
$\rho,\phi=0$. We obtain
\begin{eqnarray} &&\int\!\!\!  d3\langle{\cal D}^0(1,3){\cal
D}^0(\bar{3},2)\rangle\!=\! w(\omega; 0,0)\!\!\int\!\!\! d3\langle{\cal
D}^0(1,3)\rangle \langle{\cal D}^0(\bar{3},2)\rangle
 \nonumber \\ &&\hspace*{1.2cm}+\frac{1-w(\omega; 0,0)}{-2i\omega}
\left[
\langle{\cal D}^0\!\left(\bar{1}; 2\right)
\rangle\!  + \!
\langle{\cal D}^0\!\left(1; \bar{2}\right)\rangle
\right].
\label{dfinal}
\end{eqnarray}

\subsection{Quantum correction to the conductivity.} 
\label{sec:4.1}.

Now, we are prepared to find the correction to the conductivity.
Substituting Eqs.~(\ref{cfinal}) and (\ref{dfinal}) into
Eq.~(\ref{mainresultav}), and using Eq.~(\ref{diffaverage}) for 
$\langle {\cal D}^0\rangle$, we find
\begin{equation}
\Delta{\cal D} =
-\frac{w^2(\omega;0,0)\ln\left(\frac{1}{\omega\tau_{tr}}\right)}{4\pi^2\nu}
\frac{\nabla^2}{\left(-i\omega - D\nabla^2 \right)^2},
\label{correction1}
\end{equation} where function $w$ is given by Eq.~(\ref{wbar}). Comparing
Eq.~(\ref{correction1}) with Eq.~(\ref{diffaverage}), we see that all the
quantum correction can be
 ascribed  to the change $\Delta D$ in the diffusion constant. Restoring
the Planck constant, we obtain 
\begin{equation}
\Delta D (\omega) = -\frac{w^2(\omega;0,0)}{4\pi^2\hbar\nu}
\ln\left(\frac{1}{\omega\tau_{tr}}\right).
\label{cordiff}
\end{equation} The correction to the conductivity $\Delta\sigma$ is
related to the correction
$\Delta {\cal D}$ by Einstein relation $\Delta\sigma = se^2\nu\Delta D$,
where
$s=2$ is the spin degeneracy. We immediately find
\begin{equation}
\Delta\sigma = - \frac{e^2s}{4\pi^2\hbar}
w^2(\omega;0,0)\ln\left(\frac{1}{\omega\tau_{tr}}\right).
\label{corsigma}
\end{equation} Comparing Eq.~(\ref{corsigma}) with
Eq.~(\ref{renfunction}), we obtain the renormalization function
$\Gamma(\omega)$:
\begin{equation}
\Gamma(\omega)=w^2\left(\omega; 0, 0\right).
\label{gamma-W}
\end{equation} Here function $w$ is defined by Eq.~(\ref{wbar}).
\subsection{Universality of the weak localization correction at 
$\omega \to 0$.}
\label{sec:4.3}

The universality of weak localization correction at low frequencies,
$\Gamma (0) =1$ can be proven immediately. Indeed, function $W=1$ is a
solution of Eq.~(\ref{wequation}) and it satisfies the boundary condition
$W(\rho = a)=1$.  Because $W=1$ is the solution of nonaveraged equation in
specific disordered potential,  the averaged function $w$ also equals to
unity. Then, it follows from Eqs.~(\ref{wbar}) and (\ref{gamma-W}),  that
$\Gamma (0) =1$, which completes the proof of the universality. This fact
is well-known for the weak short range disorder, where the Born
approximation  applies. We are not aware of any proof of the universality
for the disorder of the arbitrary strength and the spatial scale.

We emphasize that the proof did not imply any small classical parameters
in the problem, and it requires only the applicability of the
semiclassical approximation, $\lambda_F \ll a, l_{tr}$. Universality is 
based on two  elements: 1) the conservation of the total number of
particles on all the spatial and time scales and 2) existence of a
diffusive motion at large spatial and time scales. Both these facts
depend neither on the strength of the scatterers nor on their spatial
size. 

It is worth mentioning also that the upper cut-off of the logarithm in
Eq.~(\ref{renfunction}) is determined by purely classical quantity
$\tau_{tr}$ and does not contain Ehrenfest time as one could expect. This
result is due to the fact that the both lower and upper limit of the
logarithm in the solution of the diffusion equation are related to the
spatial scale and not to the time scale. The upper limit of the logarithm
$\sqrt{D/\omega}$ is the typical distance at which the electron can
diffuse during time $\simeq1/\omega$. The lower linear scale is the
largest of two distances; 1) the distance between the initial and final
points, or 2) the transport mean free path -- smallest scale at which the
diffusion approximation is applicable. Because, for the problem in the
diffusive region,
 we are interested in the probability for an electron to approach its
starting point at the distance of the order of $a\lesssim l_{tr}$,  (and
by no means $\sqrt{Dt_E})$, we have to use $l_{tr}$ as the short distance
cutoff. It immediately gives $\ln (\sqrt{D/\omega}/l_{tr}) =
\ln (1/\sqrt{\omega\tau_{tr}})$.

Thus, we conclude that the weak localization correction has precisely the
same universal form as in the quantum chaos regime. However, unlike in
the QD regime, this universality persists only up to some frequency which
is much smaller than
$\tau_{tr}$ and breaks down at larger frequencies. The description of
such a breakdown is a subject of the following subsection.

\subsection{ Ehrenfest time and $\Gamma(\omega)$ at finite frequency.}
\label{sec:4.4}  Our goal now is  to find $w$ at  frequencies
$t_E^-1\lesssim\omega<\tau_{tr}^{-1}$. We would like to show that the
functional form of  $w$ is log-normal even if the parameter
$a/l_{tr}$ is not small, and derivation of the equation analogous to the
Boltzmann kinetic equation is not possible. Let us, first, neglect the
angular  diffusion in Eq.~(\ref{wequation}) at all, we will take it into
account in the end of the subsection. We rewrite Eq.~(\ref{wequation}) in
the time representation
\begin{eqnarray} &&\displaystyle{\left[\frac{\partial}{\partial t}
+\left(v_F{\bf n}\cdot
\frac{\partial}{\partial {\bf R}} - \frac{\partial U}{\partial {\bf R}}
\cdot
 \frac{\partial}
               {\partial {\bf P}}\right) -\right.}\nonumber\\
&&\displaystyle{\left.\ \left( v_F\phi\frac{\partial}{\partial \rho}-
\frac{\partial^2U}{p_F\partial R_\perp^2}
\rho\frac{\partial}{\partial \phi}\right)\right]W\left(t; {\bf n},
 {\bf R};
\phi,
\rho
\right)=0,}\nonumber\\ &&\displaystyle{\hspace{2cm}W(t) =
\int\frac{d\omega}{2\pi} e^{-2i\omega t}W(\omega),}
\label{wequationt}
\end{eqnarray} where we used the explicit form of operators
$\hat{L}_{c,r}$ from Eqs.~(\ref{L+approx}) and (\ref{L-approx}). Then, we
separate the motion of the center of mass and the relative motion of the
ends of the Cooperon. Namely, we factorize function
$W$ as
\begin{eqnarray} &&\displaystyle{W\left(t; {\bf n}, {\bf R}; \phi,
\rho\right)=
\int \frac{d{\bf R}_0d{\bf n}_0}{2\pi} W_\perp\left(t; {\bf n}_0, {\bf
R}_0; \phi,
\rho\right)\times }
\nonumber\\ &&\displaystyle{\hspace{1.2cm}\delta\left[{\bf R} -  {\bf
R}(t,{\bf R}_0, {\bf n}_0 )
\right]\delta\left[{\bf n} - {\bf n}(t,{\bf R}_0, {\bf n}_0 )
\right],}\label{wfactorized}
\end{eqnarray} where the trajectory of the center of mass ${\bf R}(t),
{\bf n}(t)$ is  found from the classical equations of motion
\begin{eqnarray} &\displaystyle{
\dot{{\bf P}}= - \frac{\partial U}{\partial {\bf R}};
 \quad \dot{{\bf R}} = \frac{{\bf P}}{m}; 
\quad {\bf n}(t)=
\frac{{\bf P}(t)}{\left|{\bf P}(t)\right|}; }&
\nonumber\\ &\displaystyle{\quad {\bf R(0)}={\bf R}_0;
\quad {\bf P(0)}={\bf n}_0p_F({\bf R}_0)}&\label{classmotion}
\end{eqnarray} and   function $W_\perp$ obeys the equation
\begin{eqnarray} &&\displaystyle{
\left[\frac{\partial}{\partial t} - v_F(t)\phi\frac{\partial} {\partial
\rho}+  F(t)
\rho\frac{\partial}{\partial \phi}\right] W_\perp=0, }\label{tildeW}\\
&&\displaystyle{v_F(t)\equiv v_F\left[{\bf R}(t,{\bf R}_0)\right], \quad
F(t)\equiv \left.\frac{\partial^2U}{p_F\partial R_\perp^2}
\right|_{{\bf R}={\bf R}(t,{\bf R}_0)} .}\nonumber
\end{eqnarray}
 
Equation (\ref{tildeW}) is invariant with respect to the scale
transformation of variables
$\rho$ and $\phi$. It invites to introduce the new variables
\begin{equation} z=\ln\sqrt{\phi^2 + \left(\frac{\rho}{a}\right)^2},
\quad
 \alpha = \arctan \frac{\phi a}{\rho}.
\label{nv}
\end{equation}

Upon this substitution, Eq.~(\ref{tildeW}) takes the form
\begin{eqnarray} &&\displaystyle{\left\{\frac{\partial}{\partial t} -
B_1(t)\sin (2\alpha)
\frac{\partial}{\partial z} + \right.}\nonumber\\ 
&&\displaystyle{\hspace{2cm}\left.\left[B_1(t) \cos (2\alpha) +B_2(t)
\right]\frac{\partial}{\partial
\alpha}\right\} W_\perp=0;} \nonumber\\ &&\displaystyle{B_{1,2}(t)=
\frac{v_F(t)}{2a} \mp
\frac{aF(t)}{2} }.
\label{tildeW1}
\end{eqnarray}

Formal solution of Eq.~(\ref{tildeW1}) is (we omit arguments ${\bf n}_0,
{\bf R}_0$ hereinafter) 
\begin{eqnarray} &&\displaystyle{W_\perp\left(t;z,\alpha\right)=
\exp \left[B_3\left(t,\alpha\right)\frac{\partial}{\partial z}\right]
W_\perp\left[0; z, \hat{\alpha}_0(\alpha,t)\right]}
\nonumber\\ &&\displaystyle{B_3\left(t,\alpha\right)\equiv
\int_0^t dt_1 B_1(t_1)\sin
\left\{2\hat{\alpha}\left[\hat{\alpha}_0
\left(\alpha,t\right),t_1\right]\right\}},
\label{formalsolution}
\end{eqnarray} where function $\hat{\alpha}(\alpha_0,t)$ satisfies the
equation of motion
\begin{equation}
\frac{\partial\hat{\alpha}}{\partial t} = B_1(t) \cos(2\hat{\alpha}) +
B_2(t),
\quad
\hat{\alpha}(\alpha_0,0) =
\alpha_0,
\label{Riccatti}
\end{equation} and function $\hat{\alpha}_0(t,\alpha)$ is implicitly
defined by the relation
\begin{equation}
\hat{\alpha}\left[\hat{\alpha}_0(t,\alpha),t\right] = \alpha.
\label{alpha0}
\end{equation} Equation (\ref{formalsolution}) enables us to find the time
evolution of function $w(t)$ from Eq.~(\ref{wbar}). Indeed, substitution
of Eq.~(\ref{wfactorized}) into Eq.~(\ref{wbar}) immediately yields
\begin{equation} w\left(t; \phi, \rho\right)=
\int \frac{d{\bf R}_0d{\bf n}_0}{2\pi S} W_\perp\left(t; {\bf n}_0, {\bf
R}_0; \phi,
\rho\right).
\label{wbarwtilde}
\end{equation} The time dependence of the function $W_\perp$ is given by
 Eq.~(\ref{formalsolution}); using this formula we obtain
\begin{equation}
\!\!w\left(t;z,\alpha\right)\!=\!\!
\int\!\! \frac{d{\bf R}_0d{\bf n}_0}{2\pi S}\!\exp \left[B_3(t)
\frac{\partial}{\partial z}\right] w\!\left[0; z,
\hat{\alpha}_0(t,\alpha)\right].
\label{formalsolutionwbar}
\end{equation}

We are interested in the time dynamics of the system at time $t$ much
larger than
$\tau_{tr}$. At such large times, function $\hat{\alpha}(\alpha_0, t)$
averaged over an arbitrary small region of  $R_0, n_0$ is a self-averaging
quantity and it no longer depends on the initial condition $\alpha_0$. 
(This fact is similar to  the randomization of the direction of momentum
in  the derivation of the diffusion equation). Therefore,  the function
$B_3$ from Eq.~(\ref{formalsolution}) becomes independent of $\alpha$.
Thus, at large times $w\left(t;z,\alpha\right)$ is also independent of
$\alpha$ and its evolution is governed by the Focker-Planck type equation:
\begin{equation}
\left[\frac{\partial}{\partial t} - {\cal
F}\left(\frac{\partial}{\partial z}\right) \right]w(t,z) = 0,
\label{FokkerPlanck}
\end{equation} where ${\cal F}(x)$ is defined as
\begin{eqnarray} &&\displaystyle{{\cal F}(x) =
\lim_{t\to\infty}\frac{1}{t}
\ln\left\{\int \frac{d{\bf R}_0d{\bf n}_0}{2\pi S}
\exp
\left[x B(t)\right]\right\},}\nonumber\\ &&\displaystyle{B(t) =
\int_0^t {dt}B_1(t)\sin
\left[ 2\hat{\alpha}(\alpha_0,t)
\right].}
\label{limit}
\end{eqnarray} In Eq.~(\ref{limit}), the initial condition $\alpha_0$ may
be chosen arbitrary. Furthermore, we will need function $w$ at large
times. In this  case $w$ is a smooth function on
$z$, and we expand ${\cal F}$ in the  Taylor series: 
\begin{eqnarray} &&\displaystyle{ {\cal F}(x)=\lambda
x+\frac{\lambda_2x^2}{2} }
\label{expansion}\\ &&\displaystyle{
\lambda = \lim_{t\to\infty}\frac{1}{t} 
\int\frac{d{\bf R}_0d{\bf n}_0}{2\pi S} B(t) }\nonumber \\
&&\displaystyle{\lambda_2 = \lim_{t\to\infty}\frac{1}{t}\left\{
\left[\int\frac{d{\bf R}_0d{\bf n}_0}{2\pi S}B^2(t)\right] - \lambda^2
t^2 \right\}} \nonumber
\end{eqnarray}

Returning to the frequency representation, we obtain the  equation
describing
 the
 drift and  diffusion of the logarithms of the coordinates:
\begin{equation}
\left[-2i\omega -
\lambda\frac{\partial}{\partial z} - 
\frac{\lambda_2}{2}\frac{\partial^2}{\partial z^2}
\right]w(\omega;z) = 0.
\label{difflog}
\end{equation} With the same accuracy, the boundary conditions  
Eq.~(\ref{WBC}) take the form
\begin{equation} w(\omega;z=0)=1
\label{BClog}
\end{equation}
 For a generic system the actual calculation of the coefficients 
$\lambda,\lambda_{2}$ can be performed, {\em e.g.} by the numerical study
of the system of equations (\ref{classmotion}) and (\ref{Riccatti}) at
times of the order of
$\tau_{tr}$ and then using Eq.~(\ref{expansion}). Analytic calculation of
coefficients $\lambda,\lambda_{2}$ requires additional model
assumptions.  Outline of such calculation  for the  weak smooth disorder
is presented in Appendix~\ref{ap:1}.

The solution of Eq.~(\ref{difflog}) at $\omega\tau_{tr} \ll 1$ and with
the boundary condition (\ref{BClog}) has the form 
\begin{equation} w(\omega;z) = \exp\left[\left(-  \frac{2i\omega}{\lambda}
+\frac{2\omega^2\lambda_2}{\lambda^3}\right)z\right].
\label{almostthere}
\end{equation} However, in order to find the renormalization function
$\Gamma (\omega)$, we need to know $w(\rho,\phi = 0)$, see
Eq.~(\ref{gamma-W}). It corresponds to taking the limit $z\to - \infty$
in Eq.~(\ref{almostthere}). One immediately realizes that $w(\rho,\phi =
0)=0$ at any finite frequency $\omega$, which would mean that the time it
takes for the quantum correction to reach its universal value is
infinite. The reason for this unphysical result lies in neglecting the
angular diffusion term in Eq. (\ref{wequationt}). It is this term that is
responsible the quantum spreading of the classical probability and it
makes the Ehrenfest time finite.

In terms of the variables (\ref{nv}), the angular diffusion operator is
given by
\begin{equation}
\frac{\partial^2}{\partial\phi^2} =\frac{1}{2}\left[ 
{e^{-2z}}\frac{\partial^2}{\partial z^2}-
\cos 2\alpha\ \frac{\partial}{\partial z} {e^{-2z}}
\frac{\partial}{\partial z}\right] + {\cal
O}\left(\frac{\partial}{\partial\alpha}\right).
\label{logangle}
\end{equation} Because function $w$ is independent of $\alpha$, we can
neglect all  the terms
${\cal O}\left(\frac{\partial}{\partial\alpha}\right)$ at all.
Furthermore, the condition $\lambda\tau_q \gg 1$ enables us to consider
the angular diffusion (\ref{logangle}) in the lowest order of
perturbation theory. As the result, Eq.~(\ref{difflog}) acquires the form
\begin{equation}
\left[2i\omega +
\lambda\frac{\partial}{\partial z} +
\frac{\lambda_2}{2}\frac{\partial^2}{\partial z^2} +
\frac{{e^{-2z}}}{2\tau_q}\frac{\partial}{\partial z}
\left(\frac{1-\gamma}{2}
\frac{\partial}{\partial z}\! + \!{\gamma} 
\right)
\right]w\! =\! 0,
\label{difflog10}
\end{equation} where the numerical coefficient $\gamma \lesssim 1$ is
given by
\[
\gamma = \lim_{t\to\infty}\frac{1}{t}\int_0^tdt 
\int\frac{d{\bf R}_0d{\bf n}_0}{2\pi S}
\cos \left[ 2\hat{\alpha}(\alpha_0,t)
\right].
\]

We now solve  Eq.~(\ref{difflog10}) with the logarithmical accuracy,
taking into account the condition $\lambda\tau_q
\gg 1$. The result is
\begin{equation} w = \exp\left[\left( \frac{i\omega}{\lambda}
-\frac{\omega^2\lambda_2}{\lambda^3}\right)
\ln\left(\frac{\lambda\tau_q}{\lambda\tau_q e^{2z} + \gamma/2}\right)
\right].
\label{sollog}
\end{equation} At $|z| \ll \ln\lambda\tau_q$,  expression (\ref{sollog})
matches with Eq.~(\ref{almostthere}).

By taking the limit $z\to -\infty$ in Eq.~(\ref{sollog}) and making use 
of Eq.~(\ref{gamma-W}),  we obtain  Eq.~(\ref{result}) with $t_E =
\frac{1}{\lambda}\ln\lambda\tau_q$. Finally, we use  estimate
(\ref{tauestem}), replace with the logarithmic accuracy
$v_F/\lambda$ to the characteristic size of the potential $a$, and arrive
to Eq.~(\ref{tE}). 

\section{Relevant perturbations.}
\label{sec:5} So far, we considered only the frequency dependence of weak
localization correction in the quantum chaos. In this section we
concentrate on two more factors which affect our results: 1) finite 
phase relaxation time
$\tau_\varphi$; 2) presence of the magnetic field;

\subsection{Effect of  finite  phase relaxation time $\tau_\varphi$} As it
was discussed in  Sec.\ref{sec:2}, the weak localization correction has
its origin in the interference between the coherent classical paths. If
the particle experiences the inelastic scattering during its motion, this
coherence is destroyed and the weak localization correction is
suppressed\cite{review,AronovAltshuler,ChakravartySchmidt}. This effect is
described conventionally by the introduction of the phase relaxation time
$\tau_\varphi$, (see Ref.~[\onlinecite{AronovAltshuler}] for a lucid
discussion of the physical meaning of $\tau_\varphi$), into the Liouville
equation for Cooperon (\ref{mainresultC}):
\begin{equation}
\left[-i\omega +\frac{1}{\tau_\varphi}+ \hat{L}_1-
\frac{1}{\tau_q}\frac{\partial^2}{\partial\phi_1^2}\right]{\cal C}
 =\delta_{12}.
\label{KC1phi}
\end{equation} The  equation for the diffuson (\ref{mainresultD})  remains
unchanged as well as the  relations (\ref{mainresult0}) and
(\ref{mainresult}) between the correction to the classical probability
and the Cooperon and diffusons. 

Thus, we have to modify the Cooperon part of Eq.~(\ref{corsigma}); namely 
Eq.~(\ref{solution1}) acquires the form
\begin{equation} { C}\left(\phi,
\rho\right) = \frac{w\left(\omega+{i}/{\tau_\varphi}; \phi, \rho
\right)}{4\pi D} 
\ln
\left(\frac{\tau_{tr}^{-1}}{\sqrt{\omega^2+\tau_\varphi^{-2}}}\right).
\label{solutionphi}
\end{equation}

Comparing Eqs.~(\ref{cfinal}) and (\ref{solutionphi}), we obtain with the
help of Eqs.~(\ref{corsigma}) and (\ref{gamma-W})
\begin{equation}
\Delta\sigma\! =\! - \frac{e^2s}{4\pi^2\hbar} \left[
\Gamma\left(\omega\right)
\Gamma\left(\omega+\frac{i}{\tau_\varphi}\right)
\right]^{1/2}
\!\ln
\left(
\frac{\tau_{tr}^{-1}}{\sqrt{\omega^2+\tau_\varphi^{-2}}}
\right).
\label{resphi0}
\end{equation}  For $\omega =0$ and $\tau_\varphi \gg \tau_{tr}$,
expression (\ref{resphi0}) acquires the form
\begin{equation}
\Delta\sigma =-
\frac{e^2s}{4\pi^2\hbar}\exp\left[-\frac{t_E}{\tau_\varphi}
\left( 1-\frac{\lambda_2}{\lambda^2\tau_\varphi}
\right)\right]
\ln\left(\frac{\tau_\varphi}{\tau_{tr}}\right).
\label{resphi}
\end{equation}
 The factor $e^{-t_E/\tau_\varphi}$ in Eq.~(\ref{resphi}) can be easily
understood.  A relevant trajectory may close not earlier than it leaves
the Lyapunov region; factor
$e^{-t_E/\tau_\varphi}$ is nothing but the probability for an  electron
not to be scattered inelastically  while it is in the Lyapunov region.
Let us notice also that the dependence of the weak localization
correction  on the phase relaxation time is always slower  than an
exponential.  The reason for this is the following. The probability for a
trajectory to leave the Lyapunov region during time interval
$\tau_\varphi/2$ is determined by the corresponding Lyapunov exponent and,
thus,  it can be increased due to the fluctuation of this exponent. The
probability to find such a fluctuation is given by the Gaussian
distribution. The optimization of the  product of these two probabilities
immediately yields the exponential  factor in Eq.~(\ref{resphi}).

At this point, we should caution the reader, that the fact that the same
$\tau_\varphi$ enters into the logarithmic factor and into the
renormalization factor $\Gamma$ in Eq.~(\ref{resphi0})  is somewhat model
dependent. Strictly speaking, this statement is valid only if the phase
breaking occurs via single inelastic process with the large energy
transfer. If the main mechanism of the phase breaking is associated with
the large number of scattering events with the small energy
transfer\cite{AronovAltshuler,ChakravartySchmidt}, the phase breaking
occurs when the distance $\rho$ between the Cooperon ends is large
enough, 
$\sqrt{D/T} \lesssim \rho \lesssim \sqrt{D\tau_\varphi}$. Thus, this
mechanism does not affect the Cooperon in the Lyapunov region at all.
Further discussion of the microscopic mechanisms of the phase breaking is
beyond the scope of the present paper.

\subsection{Effect of magnetic field}

Similar to the phase relaxation time, the effect of the magnetic field on
the weak localization correction is taken into account by the change in
the equation of motion for the Cooperon
only:\cite{review,AronovAltshuler,HLN}
\begin{equation}
\left[-i\omega + \hat{L}_1 +\frac{2ie}{c}{\bf v}_1{\bf A}_1 -
\frac{1}{\tau_q}\frac{\partial^2}{\partial\phi_1^2}
\right]{\cal C}
 =
\delta_{12},
\label{KC1H}
\end{equation} where ${\bf A}_1={\bf A}({\bf R}_1)$ is the vector
potential of the external magnetic field. Cooperon given by
Eq.~(\ref{KC1H}) is not a gauge invariant quantity but
${\cal C}(1,\bar{1})$ is. It is very convenient to separate the gauge
noninvariant part of the Cooperon explicitly by writing
\begin{equation} {\cal C} = \exp\left(\frac{2ie}{c}\int{\bf A}d{\bf
r}\right) {\cal C}_{\rm gi},
\label{giCooperon}
\end{equation} where integration in the first factor is carried out along
the straight line connecting the Cooperon ends. Substituting
Eq.~(\ref{giCooperon}) into Eq.~(\ref{KC1H}), we obtain  the gauge
invariant part of the Cooperon
\begin{equation}
\left[-i\omega + \hat{L}_1 +
\frac{i\left[{\bf z}\times
\mbox{\boldmath $r$}
\right]}{\lambda_H^2}\right]{\cal C}_{\rm gi}
 =
\delta_{12},
\label{KC1Hgi}
\end{equation} where ${\bf r} = {\bf R}_1 - {\bf R}_1$  and
$\lambda_H=\left(c\hbar/eH\right)^{1/2}$ is the magnetic length. When the
ends of the Cooperon coincide, ${\cal C}_{\rm gi} = {\cal C}$, and,
therefore,  the correction to the conductivity (\ref{corsigma}) is
modified as
\begin{equation}
\Delta\sigma = -\frac{se^2}{\pi\hbar} w(\omega;0,0) \left( \langle {\cal
C}_{\rm gi}\left(1, \bar{1}\right)\rangle D\right).
\label{corsigmam}
\end{equation}

Our purpose now is to obtain the expression for ${\cal C}_{\rm gi}$.
Similar to the case of zero magnetic field, we would like to separate the
problem into Lyapunov and diffusion region. This separation, however, is
valid only if the condition
\begin{equation}
\lambda_H \gg l_{tr}
\label{condition}
\end{equation} holds. This condition follows from the fact that, the
characteristic area enclosed by the relevant trajectory should not exceed
$\lambda_H^2$. If Eq.~(\ref{condition}) is not fulfilled, the trajectory
should turn back at the distances  much smaller than
$l_{tr}$.  The probability of such an event  is determined by the optimal
configurations consisting of a small number of scatterers and, thus,
separation of the diffusion region is not
possible\cite{KawabataDyakonov}. In all the subsequent calculations, we
assume that the condition (\ref{condition}) is met.

In the diffusion region, the Cooperon satisfies the equation
\begin{equation}
\left[-i\omega -  D\left(\mbox{\boldmath
$\nabla$}_\rho+\frac{i\left[{\bf z}\times
\mbox{\boldmath $\rho$}
\right]}{\lambda_H^2}\right)^2\right]\langle {\cal C}_{\rm gi}\rangle = 
\delta \left(\mbox{\boldmath $\rho$}\right).
\label{mfdiffregion} 
\end{equation}

At $a\lesssim\rho \lesssim l_{tr}$, the Cooperon
${\cal C}_{\rm gi}$ ceases to depend on $\rho$, and we have with the
logarithmic accuracy
\begin{equation}
\langle {\cal C}_{\rm gi}\rangle \approx
\frac{1}{4\pi D}
\left[\ln
\left(\frac{1}{\omega\tau_{tr}}\right)-{Y}\left(\frac{
D}{-i\omega\lambda_H^2}\right)\right],
\label{mfdiffregionb} 
\end{equation} where dimensionless function ${Y}\left(x\right)$ is given
by\cite{HLN}
\begin{equation} {Y}\left(x\right) =
\Psi\left(\frac{1}{2}+\frac{1}{4x}\right) + \ln 4x,
\label{Y}
\end{equation} and $\Psi (x)$ is the digamma function.

The solution in the Lyapunov region with the boundary condition
(\ref{mfdiffregionb}) can be represented in a form similar to
Eq.~(\ref{solution1}) 
\begin{equation}
\langle {\cal C}_{\rm gi}\rangle = \frac{w_c(\omega; \phi,
\rho)} {4\pi D}
\left[\ln
\left(\frac{1}{\omega\tau_{tr}}\right)-{Y}\left(\frac{
D}{-i\omega\lambda_H^2}\right)\right].\label{msolution1} 
\end{equation} Here, function $w_c$ is  related to $W_c$ by
Eq.~(\ref{wbar}), however, the equation for the latter function, see
Eq.~(\ref{wequationt}), is modified:
\begin{eqnarray} &&\displaystyle{\!\!\!\left\{\frac{\partial}{\partial t}
+\left[v_F{\bf n}
\cdot
\frac{\partial}{\partial {\bf R}} - \frac{\partial U}{\partial {\bf R}}
\cdot
 \frac{\partial}
               {\partial {\bf P}}\right]-\right.}\label{mwequationt}\\
&&\displaystyle{\left.\ \left[ v_F\left(\phi\frac{\partial}{\partial
\rho} + \frac{2i\rho} {\lambda_H^2}
\right)-
\frac{\partial^2U}{p_F\partial R_\perp^2}
\rho\frac{\partial}{\partial \phi}\right]\right\}W_c=0,}.
\nonumber
\end{eqnarray} 
Equation (\ref{mwequationt}) is supplied with the boundary
condition (\ref{WBC}).

Now, we will show that this modification does not affect  function $W_c$
in the Lyapunov region  provided that condition (\ref{condition}) holds.
Thus, the renormalization function
$\Gamma (\omega)$ is not affected by the magnetic field. In order to
demonstrate this we use the following arguments. The effect of the extra
in comparison with Eq.~(\ref{wequationt}) term in 
Eq.~(\ref{mwequationt}) can be taken into account by multiplying function
$W_\perp$  from Eq.~(\ref{wfactorized}) by the factor $\exp\left(2i{\cal
A}(t)/\lambda_H^2\right)$, where ${\cal A}(t)$ is the  area enclosed by
the trajectory in the Lyapunov region and it is given by
\begin{equation} {\cal A}(t)=\int_0^t dt_1v_F(t_1)\rho(t_1).
\label{A}
\end{equation} Let us estimate the maximal value of area ${\cal A}$. In
the Lyapunov region, the distance $\rho$ does not exceed the
characteristic scale of the potential $a$. In the vicinity of the
boundary of the Lyapunov region
$\rho$ depends exponentially on time
$\rho (t) \simeq a e^{\lambda t}$ (here time
$t<0$ is counted from the moment of arrival of the trajectory to the
boundary of the Lyapunov region). Substituting  this estimate  into
Eq.~(\ref{A}), we obtain
\begin{equation} {\rm max}|{\cal A}| \simeq av_F/\lambda \lesssim
l_{tr}^2.
\label{Aest}
\end{equation} Comparing  estimate (\ref{Aest}) with the condition
(\ref{condition}), we conclude that
$|{\cal A}| \ll \lambda_H^2$ and, therefore, the magnetic field has no
effect in the Lyapunov region.

Thus, final formula for the weak localization correction in the magnetic
field $H$ reads
\begin{equation}
\Delta\sigma(H,\omega) - \Delta\sigma(0,\omega)=
\frac{e^2s}{4\pi^2\hbar}\Gamma(\omega){Y}\left(\frac{
D}{-i\omega\lambda_H^2}\right),
\label{mresult}
\end{equation} where functions $\Gamma$ and $Y$ are defined by
Eqs.~(\ref{result}) and (\ref{Y}) respectively. It is worth noticing that
the effects of the phase relaxation, see Eq.~(\ref{resphi}), and of the
magnetic field on the renormalization function are different. This is
because the effect of the phase relaxation is determined by the time the
particle spends in the Lyapunov region, which is significantly larger than
$\tau_{tr}$, whereas the effect of the magnetic field is governed by the
area enclosed by the trajectory in the Lyapunov region which is always
much smaller than
$l_{tr}^2$.

For the weak magnetic fields, $\lambda_H^2 \gg D/{\rm max}(\omega,
\tau_\varphi^{-1})
$, we obtain from Eq.~(\ref{mresult}) 
\begin{eqnarray} &&\Delta\sigma(H) -\!
\Delta\sigma(0)=\!\label{lowfields} \\
&&\hspace*{1.0cm}\frac{e^2s}{6\pi^2\hbar}
\left[
\Gamma\left(\omega\right)
\Gamma\left(\omega +
\frac{i}{\tau_\varphi}\right)
\right]^{1/2}
\left[\frac{
D\tau_\varphi}{\left(1-i\omega\tau_\varphi\right)\lambda_H^2}\right]^2.
\nonumber
\end{eqnarray} The study of the frequency dependence or temperature (via
$\tau_\varphi$) of the magnetoresistance may provide an additional tool
for measuring the Lyapunov exponent.

\section{Weak localization in the ballistic cavities}
\label{sec:6}

In this section we study how the Lyapunov region affects  the weak
localization correction in the ballistic cavities. At zero-frequency and
$\tau_\varphi \to
\infty$ this problem was studied in
Refs.~[\onlinecite{Stone,Mello,Argaman}].

For the sake of simplicity, we restrict ourselves to the case of zero
magnetic field 
$H =0$ and concentrate upon the dependence  of the weak localization
correction to the conductance $\Delta g$ of a ballistic cavity on
frequency
$\omega$ and phase relaxation time $\tau_\varphi$. The effect of the
magnetic field on the weak localization was studied in
Ref.~[~\onlinecite{Stone}~].

{\narrowtext 
\begin{figure}[h]
\ifthenelse{\equal{\showfigures}{yes}} {
\hspace*{0.1cm}\psfig{figure=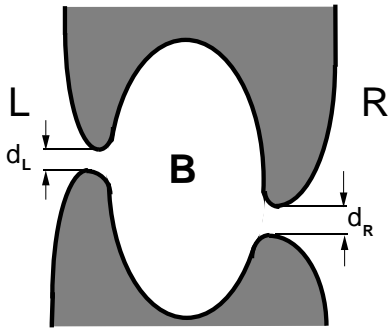,height=2.5in} } {\vspace*{2.5in}}
\vspace{0.4cm}
\caption{ Schematic view of the ballistic cavity ``B'' between two
reservoirs ``L'' and ``R''.  }
\label{fig:3}
\end{figure} }

Let us consider the system consisting of three cavities, see
Fig.~\ref{fig:3}, connected by channels. The size of the central cavity
(``B'' in Fig.~\ref{fig:3}) is much smaller than that of the outer
cavities (``L'' and ``R'' in Fig.~\ref{fig:3}) which act as reservoirs. 
The conductance of the system is  controlled by the channels so that
their widths $d_{L,R}$ are much smaller than the characteristic size of
the central cavity, $d_{L,R} \ll a$. We assume that the motion of an
electron in the channel still can be described by the classical Liouville
equation, which implies $d_{L,R} \gg
\lambda_F$. 

Because of the inequality $d_{L,R} \ll a$, the time it takes
 to establish the equilibrium distribution function  in the cavity is much
smaller than the escape time. (The equilibration time is of the order of
the flying time of the electron across the cavity.) Under such conditions
the classical escape times  from the cavity through the left (right)
channel
$\tau_{L(R)}$ are given by 
\begin{equation}
\!\!\frac{1}{\tau_{L(R)}}\!=\!
\frac{1}{{\cal A}_B}\!\int\!\! \frac{dn}{2\pi}\!\!
\int\!\! \theta\left({\bf n}\! \cdot\! d\mbox{\boldmath
$\ell$}\right) v_{L(R)}{\bf n}\!  \cdot\!  d\mbox{\boldmath
$\ell^{L(R)}$}\! =\! \frac{d_{L(R)}v_{L(R)}}{2{\cal A}_B}, 
\label{taus}
\end{equation} where ${\cal A}_B$ is the area of the cavity, the linear
integration is performed along the narrowest cross-section of the 
corresponding channel,
$d\mbox{\boldmath
$\ell$}^{L(R)}$ is directed outside the cavity  ``B'' normal to the
integration line, and
$v_{L(R)}$ are the Fermi velocities in the contacts. Equation (\ref{taus})
corresponds to the classical Sharvin formula\cite{Sharvinconductance} for
2D case and the escape times are related to the classical conductance of a
single channel $g_{L(R)}$ by
\begin{equation} g_{L(R)} = \frac{se^2\nu {\cal A}_B}{\tau_{L(R)}}. 
\label{taug}
\end{equation}
 
If the external bias $eV(t)$ is applied to, say, left reservoir,  (the
right reservoir is maintained at zero bias), the electric current
$I$ from the left to the right reservoir appears. This current is  linear
in the applied bias:
\begin{eqnarray} &\displaystyle{I(t) \equiv - \dot{Q}_L(t) =
\int_{-\infty}^tdt^\prime g(t-t^\prime)V(t^\prime),}&\nonumber\\
&\displaystyle{ g(t) = \int\frac{d\omega}{2\pi}e^{-i\omega t}g(\omega),}&
\label{gsystem}
\end{eqnarray} where ${Q}_L$ is the charge of the left reservoir. Relation
Eq.~(\ref{gsystem}) defines the conductance of the system $g(\omega)$.
Performing actual calculations in Eq.~(\ref{gsystem}), one has to take
into account the condition of the electroneutrality in the cavity ``B'',
$\dot{Q}_B = 0$.  The electroneutrality requirement is valid at times
larger the characteristic time  of the charge relaxation. This time
$\tau_c$ can be estimated as $\tau_c
\simeq C_B/({\rm max}\ g_{L,R})$, where $C_B$ is the capacitance of the
cavity. Using estimate $C_B
\sim a$ and formulas (\ref{taus}), (\ref{taug}), we find $\tau_c \simeq
\tau_{fl} a_B/ ({\rm max}\ d_{L,R})$, where  $\tau_{fl} = a/v_F$ is the
flying time of the electron across the cavity, and $a_B$ is the screening
radius in 2D electron systems. For  wide channels $d_{L,R} \gg \lambda_F
\simeq a_B$, we have
$\tau_c\ll\tau_{fl}$.  We are interested in the dynamics of the system at
time much larger than the flying time and, therefore,  we can assume that
the electroneutrality holds.

Then, the standard linear response calculations enable us to relate the
conductance
$g$ to the diffuson ${\cal D}$ defined in Sec.~\ref{sec:2}. The charge
response  in $i$th cavity, $Q_i$ to the applied biases
$V(t),V_B(t) = V,V_Be^{-i\omega t}$ can be expressed by means of the
polarization operator as
\begin{equation} Q_i =\! e^2\!\!\int\!\! d{\bf r}_1d{\bf
r}_2\Pi\left(\omega; {\bf r}_1,{\bf r}_2
\right)\!\theta_i({\bf r_1})\!
\left[V\theta_L({\bf r_2}) + V_B\theta_B({\bf r_2})
\right],
\label{cavityprom}
\end{equation} where function
$\theta_i({\bf R})$   equals to unity if vector
${\bf R}$ lies in the $i$th region ($i=L,R,B$) and equals to zero
otherwise. The potential $V_B$ is to be found self-consistently from the
electroneutrality requirement. Substituting Eq.~(\ref{polarization}) into
Eq.~(\ref{cavityprom}) and making use of Eqs.~(\ref{mixedrepD}), and
(\ref{diffuson1}), we obtain with the help of definition (\ref{gsystem})
\begin{eqnarray} &&\displaystyle{g(\omega) = se^2\nu\left\{-i\omega {\cal
A}_L  +\omega^2\left[{\cal D}_{LL}(\omega) + {\cal
D}_{LB}(\omega)\frac{V_B}{V}\right]\right\};} 
\nonumber\\ &&\displaystyle{{\cal D}_{ij}(\omega)
 \equiv \int\frac{d{\bf n_1}d{\bf n_2}d{\bf R_1}d{\bf
R_2}}{\left(2\pi\right)^2}
\theta_i({\bf R_1})\theta_j({\bf R_2})}\nonumber\\
&&\displaystyle{\hspace*{3cm}\times{\cal D}_{\epsilon_F} (\omega;{\bf
n_1},{\bf R_1};{\bf n_2},{\bf R_2}),}
\label{gd}
\end{eqnarray} where ${\cal A}_i$ is the area of the corresponding region
($i=L,R,B$). 

The electroneutrality condition, $Q_B=0$, gives us the equation for the
potential of the cavity $V_B$. Using  Eq.~(\ref{cavityprom}) for $i=B$,
we find with the help of Eqs.~(\ref{polarization}), (\ref{mixedrepD}) and
(\ref{diffuson1}) 
\begin{equation} i\omega{\cal D}_{LB}(\omega)V+
\left[ {\cal A}_B +i\omega{\cal D}_{BB}(\omega)
\right]V_B=0.
\label{neutrality} 
\end{equation}

\subsection{Classical conductance}

Let us first calculate the classical conductance $g_{cl}$ of the system. 
We consider the frequencies $\omega$, much smaller than the flying time
of the electron in a cavity. Assuming that the motion in the cavity is
ergodic and the areas of the reservoirs are large, ${\cal A}_{L(R)}/{\cal
A}_B \gg
\omega\tau_{L(R)}$, we obtain that  the diffuson changes only within the
channels. For ${\cal D}_{ij}$ from Eq.~(\ref{gd}) we  find
\begin{mathletters}
\begin{eqnarray} &\displaystyle{{\cal D}_{BB}^0(\omega)= 
\frac{{\cal A}_B}{-i\omega +\frac{1}{\tau_B}}, \quad \frac{1}{\tau_B}=
\frac{1}{\tau_L}+\frac{1}{\tau_R}}&
\label{Dbbclassical}\\ &\displaystyle{{\cal D}_{jj}^0(\omega)=
\frac{{\cal A}_j}{-i\omega} +
\frac{{\cal A}_B}{\tau_j\omega^2}  -
\frac{{D}_{BB}^0}{\left(\tau_j\omega\right)^2}, \quad j=L,R  };&
\label{Dllclassical}\\ &\displaystyle{{\cal D}_{jB}^0(\omega)= {\cal
D}_{Bj}^0(\omega)= 
\frac{{\cal D}_{BB}^0(\omega)} {-i\omega\tau_j},  \quad j=L,R};&
\label{Dlbclassical}\\ &\displaystyle{{\cal D}_{LR}^0(\omega)= {\cal
D}_{RL}^0(\omega)=  -\frac{ {\cal D}_{BB}^0(\omega)
     } {\omega^2\tau_L\tau_R}. }&
\label{Dlrclassical}
\end{eqnarray}
\label{Dcl}
\end{mathletters} Equation (\ref{Dbbclassical}) describes the
exponentially decaying in time probability to find the electron in the
cavity ``B'' if it started in this cavity. First term in
Eq.~(\ref{Dllclassical}) corresponds to the classical correlator of the
$j$th reservoir disconnected from the cavity,  the second term describes
the finite probability for the electron to enter cavity ``B'' from $j$th
reservoir, and the third term corresponds to the process in which an
electron from $j$th reservoir visits the cavity once and then comes back.
Equation~(\ref{Dlbclassical}) gives the probability for the electron to
appear in the $j$th reservoir starting from the cavity. Finally,
Eq.~(\ref{Dlrclassical}) is the probability for an electron to get from
the left to the right reservoir.

Substituting Eqs.~(\ref{Dcl})  into Eq.~(\ref{neutrality}),  we find that
the bias of the cavity $V_B$ does not depend on frequency,
$V_B= g_L/\left(g_L+g_R\right)$. Then, by substitution  Eqs.~(\ref{Dcl})
into Eqs.~(\ref{gd}), we  obtain with the help of Eq.~(\ref{taug}) 
\begin{equation} g_{cl} = \frac{g_Lg_R}{g_L+g_R}
\label{gclassical}
\end{equation} in agreement with the Kirchhoff law. It is worth
mentioning that the result (\ref{gclassical}) at
$\omega=0$ can be obtained without requirement of the electroneutrality. 

\subsection{Weak localization correction}

In order to calculate the weak localization correction to the conductance
$\Delta g(\omega)$ we have to find the  correction to the classical
correlator 
$\Delta {\cal D}$ and then use Eq.~(\ref{gd}) and (\ref{neutrality}). For
such calculation, it is most convenient to use Eq.~(\ref{mainresult}).
Our strategy will be analogous to the one we used in Sec.~\ref{sec:4} for
the calculation of the correction to the conductivity.

Integrating both sides of Eq.~(\ref{mainresult}) over the coordinates $1$
and $2$ within the regions specified by $\theta$-functions in
Eq.~(\ref{gd}) and using obvious relation ${\cal D}(1,2) = {\cal
D}(\bar{2}, \bar{1})$, we obtain  
\begin{eqnarray} &&\Delta{\cal D}_{ij} = \Delta{\cal D}_{ji}^{(1)} + 
\Delta{\cal D}_{ji}^{(2)}, \label{mainresultcavity}\\ &&\Delta{\cal
D}_{ji}^{(1)}=\int d1 
\left[
\raisebox{0mm}[2mm][2mm]{$
				{\cal D}^0_j(1)\theta_i({\bf R}_1) +
				{\cal D}^0_i(\bar{1})\theta_j({\bf R}_1)
                         $}
\right]
\frac{{\cal C}^0(1,\bar{1})}{2\pi\nu},\nonumber\\ &&\Delta{\cal
D}_{ji}^{(2)}=\int d1 
 {\cal D}^0_i(1){\cal D}^0_j(1)
\left[ 2i\omega-
\hat{L}_1 +\frac{1}{\tau_q}\frac{\partial^2}{\partial\phi^2_1}
\right]
\frac{{\cal C}^0(1,\bar{1})}{2\pi\nu},
\nonumber
\end{eqnarray} where
\begin{equation} {\cal D}^0_i(1)\equiv\int d2 \theta_i({\bf R}_2){\cal
D}^0(2,1).
\label{Dj}
\end{equation} Here, we use the short hand notation $l\equiv ({\bf n}_l,
{\bf R}_l)$,  integration over the  phase space on the energy shell is
defined as 
$dl \equiv d{\bf n}_ld{\bf R}_l/2\pi$, and the time reversed coordinate
$\bar{l}$ is given by
$\bar{l}\equiv (-{\bf n}_l, {\bf R}_l)$.

It is noteworthy, that the quantum correction $\Delta {\cal D}_{ij}$
satisfy the charge conservation condition
\begin{equation}
\sum_{j=L,R,B}\Delta {\cal D}_{ij} = 0,\quad i=L,R,B
\label{chconserv}
\end{equation} which can be easily proven with the help of the relation
$\sum_i{\cal D}^0_i(1) = \frac{1}{-i\omega}$ and Eq.~(\ref{mainresultD}).
Equation (\ref{chconserv}) enables us to consider only non-diagonal
elements of
$\Delta {\cal D}_{ij}$ which is technically easier.

Analogous to the discussion in Sec.\ref{sec:4}, we assume that the
Cooperon part of the expression can be calculated independently of the
diffuson part. This is because the classical trajectories corresponding
to these quantities traverse essentially the different regions of the
phase space. 

First, we use this assumption to evaluate contribution $\Delta {\cal
D}^{(1)}$ from Eq.~(\ref{mainresultcavity}). We notice that the classical
trajectory can close only inside the cavity. Therefore, the Cooperon
${\cal C}(1,\bar{1})$ also exists only inside the cavity ``B''. For the
calculation of the diffuson, we notice that at times much larger than the
flying time across the cavity $\tau_{fl}$ the position of the electron and
its momentum is randomized. It suggests to use the approximation
\begin{equation} {\cal D}^0_i(1) \approx \frac{1}{{\cal A}_B}\int d1{\cal
D}^0_i(1)\theta_B({\bf R}_1) = \frac{{\cal D}^0_{iB}}{{\cal A}_B},
\label{approx1}
\end{equation} if vector $R_1$ lies inside the cavity. Here, function
${\cal D}^0_{ij}$ is defined in Eq.~(\ref{Dcl}). Using
Eq.~(\ref{approx1}), we obtain
\begin{equation}
\Delta{\cal D}_{ji}^{(1)}= 
\left[
\raisebox{0mm}[2mm][2mm]{$
				{\cal D}^0_{jB}\delta_{iB} +
				{\cal D}^0_{iB}\delta_{jB}
                         $}
\right]
\frac{\langle {\cal C}^0(1,\bar{1})\rangle }{2\pi\nu}
\label{deltaD1}
\end{equation} where the average inside the cavity is defined as
\[
\langle\dots \rangle = \frac{1}{{\cal A}_B}\int d1 \theta_B({\bf R}_1)
\dots\ .
\] Let us turn to the calculation of the contribution $\Delta{\cal
D}^{(2)}_{ij}$. As we already saw in Secs.~\ref{sec:2} and \ref{sec:4},
 two diffusons can not be averaged independently, because motion of their
ends are governed by the same potential during period $t_E\gg \tau_{tr}$.
On the other hand, the randomization of the motion of the center of mass
occurs during in a time interval of the order $\tau_{fl}$. Therefore, we
can approximate
\begin{equation} {\cal D}^0_i(1){\cal D}^0_j(1) \approx \theta_B({\bf
R}_1)\langle{\cal D}^0_i(1){\cal D}^0_j(1) \rangle.
\label{approx2}
\end{equation} Expression (\ref{approx2}) is written in the lowest order
in small
 parameter ${\cal A}_B/{\cal A}_{L,R}$, and we exclude from our
consideration cases $i=j=L,R$. In the latter cases, there are also
non-vanishing contributions in Eq.~(\ref{approx2}) corresponding to the
coordinate ${\bf R}_1$ in the reservoirs $L$ or $R$. This would require
more careful investigation of the behavior of the diffuson in the
channels. We, however, simply bypass this difficulty by utilizing identity
(\ref{chconserv}) for the calculation of the diagonal elements
$\Delta{\cal D}_{LL}$ and $\Delta{\cal D}_{RR}$.  Using the approximation
(\ref{approx2}), we find
\begin{equation}
\Delta{\cal D}_{ji}^{(2)}=
\langle {\cal D}^0_i(1){\cal D}^0_j(1)\rangle\!
\int\!d1\ \theta_B({\bf R}_1)
 \left[\raisebox{0mm}[2mm][2mm]{$ 2i\omega-
\hat{L}_1 
$}
\right]
\frac{{\cal C}^0(1,\bar{1})}{2\pi\nu}.
\label{d2}
\end{equation} Liouvillean operator $\hat{L}_1$ in the second term of the
RHS of Eq.~(\ref{d2}) is the total derivative along a classical
trajectory and, therefore, it  can be reduced to the linear integrals
across the channels
\begin{eqnarray} &&\int\!d1\ \theta_B({\bf R}_1)\hat{L}_1 {\cal
C}^0(1,\bar{1}) =\label{d3} \\ &&\hspace*{1.7cm} \int
\frac{dn_1}{2\pi}\left(
\int\! {\bf n}_1 \cdot d\mbox{\boldmath $\ell^{L}_1$} +\!\!
\int\! {\bf n}_1 \cdot d\mbox{\boldmath $\ell^{R}_1$} 
\right)  v_F {\cal C}^0(1,\bar{1}) 
\nonumber
\end{eqnarray} where the linear integration is defined similar to that in
Eq.~(\ref{taus}). Then, we notice that a classical trajectory can close 
only inside the cavity. Therefore, only the Cooperons  with the initial
momentum directed inside the cavity exist. Assuming that the
randomization of the momentum direction occurs only inside the cavity and
considering the times much larger than the flying time,  we conclude that
the Cooperon in the contact  vanishes if its moment
${\bf n}_1$  directed inside the cavity and for the moment directed
outside the  cavity Cooperon coincides with its value inside the cavity,
${\cal C}(1,\bar{1})=\theta\left({\bf n}_1\cdot d{\bf
\ell}_{L(R)}\right)\langle {\cal C}(1,\bar{1})\rangle$ for the coordinate
$R_1$ located in the left or right channel respectively. This enables us
to reduce Eq.~(\ref{d3}) to the simple form
\begin{equation}
\int\!d1\ \theta_B({\bf R}_1)\hat{L}_1 {\cal C}^0(1,\bar{1}) = 
\frac{{\cal A}_B}{\tau_B}\langle {\cal C}(1,\bar{1})\rangle,
\label{sd3} 
\end{equation} where the total escape time $\tau_B$ is defined in
Eq.~(\ref{Dbbclassical}). Deriving Eq.~(\ref{sd3}), we use the definition
of the escape times (\ref{taus}).  Arguments above are essentially
equivalent to those in the derivation of the classical Sharvin
conductance\cite{Sharvinconductance}.

Combining formulas (\ref{deltaD1}), (\ref{d2}), (\ref{sd3}) and
(\ref{mainresultcavity}), we obtain
\begin{eqnarray} &&\Delta{\cal D}_{ji} =
\frac{\langle {\cal C}^0(1,\bar{1})\rangle }{2\pi\nu}
\left[\raisebox{0mm}[3mm][3mm]{$
		{\cal D}^0_{jB}\delta_{iB} +
				{\cal D}^0_{iB}\delta_{jB} +$}
\right.
\label{dd10} \\ &&\quad\quad\quad\quad \left. 
\left(2i\omega-\frac{1}{\tau_B}\right) {\cal A}_B
\langle {\cal D}^0_i(1){\cal D}^0_j(1) \rangle \right].
\nonumber
\end{eqnarray} We reiterate that Eq.~(\ref{dd10}) is not applicable for
the case of $i=j =L,R$. In order to find the diagonal elements
$\Delta{\cal D}_{LL}$ and $\Delta{\cal D}_{RR}$, one has to use the
identity (\ref{chconserv}).

The calculation of the corresponding averages $\langle {\cal
C}^0(1,\bar{1})\rangle$ and $\langle {\cal D}^0_i(1){\cal D}^0_j(1)
\rangle$ is performed along the lines of the derivations in
Sec.~\ref{sec:4}. In the calculation of the Cooperon the only change is in
the expression (\ref{diffregion}) for the Cooperon outside the Lyapunov
region
\begin{equation} C(\phi,\rho) = \frac{1}{{\cal A}_B \left(-i\omega + 
\frac{1}{\tau_B}\right)},
\label{cooperondiffb}
\end{equation} which is analogous to Eq.~(\ref{Dbbclassical}) The
solution for the Cooperon in the Lyapunov region is analogous to one
presented in Secs.~\ref{sec:4.2} and  \ref{sec:4.4}. The calculation of
function $w$ may be performed for the cavity disconnected from the
reservoirs, provided that the condition $t_E \ll \tau_B$ holds. As the
result we obtain
\begin{equation}
\langle {\cal C}(1,\bar{1})\rangle = 
\frac{ w(\omega,0,0) } { {\cal A}_B
\left(-i\omega + 
\frac{1}{\tau_B}
\right) }.
\label{cbresult}
\end{equation} In the calculation of the product of two diffusons 
$\langle {\cal D}^0_i(1){\cal D}^0_j(1)\rangle$ the change should be made 
in Eq.~(\ref{KD7}). The reason for this is that the integration over ${\bf
R}, {\bf n}$ for the reducing Eq.~(\ref{KD5}) to Eq.~(\ref{KD6}) is
performed now only inside the cavity. As a result, one more term
\[
\int \frac{d{\bf n}d{\bf R}}{2\pi}
\theta_B({\bf R})\hat{L}_c {\cal M}_1
\left(1,2;{\bf n},{\bf R};\phi,\rho\right) \approx
\frac{{ M}_1 \left(1,2;\phi,\rho\right)}{\tau_B}
\] [cf. with the derivation of Eq.~(\ref{sd3})] has to be added to the
LHS of Eq.~(\ref{KD6}). Equation (\ref{KD7}), then, acquires the form 
\[ M_1(1,2;\rho, \phi) = \frac{\langle{\cal D}^0\!\left(\bar{1}; 2\right)
\rangle\!  + \!
\langle{\cal D}^0\!\left(1;
\bar{2}\right)\rangle}{-2i\omega+\frac{1}{\tau_B}},
\] and we obtain instead of Eq.~(\ref{dfinal})
\begin{eqnarray} &&{\cal A}_B \langle{\cal D}^0_i(1){\cal
D}^0_j(1)\rangle\!=\! w(\omega; 0,0)\frac{{\cal D}^0_{iB}{\cal D}^0_{jB}}
{{\cal A}_B}+
 \nonumber \\ &&\hspace*{1.2cm}\frac{1-w(\omega;
0,0)}{-2i\omega+\frac{1}{\tau_B}}
\left[{\cal D}^0_{iB}\delta_{jB} + {\cal D}^0_{jB}\delta_{iB},
\right],
\label{dbfinal}
\end{eqnarray} where functions ${\cal D}^0_{iB}$ are given by
Eqs.~(\ref{Dbbclassical}) and (\ref{Dlbclassical}). Result
(\ref{dbfinal}) is not applicable for
$i=j=L,R$ cases. Deriving Eq.~(\ref{dbfinal}), we used
Eq.~(\ref{approx1}) for the average of the single diffuson $\langle {\cal
D}^0_i(1)\rangle$.

\begin{mathletters}
\label{corr10} Substituting Eqs.~(\ref{cbresult}) and (\ref{dbfinal}) into
Eq.~(\ref{dd10}), we obtain with the help of Eqs.~(\ref{Dcl}) and
(\ref{gamma-W})
\begin{eqnarray}
\Delta{\cal D}_{BB}\left(\omega\right) = 
\frac{\Gamma(\omega)}{2\pi\nu}
\frac{\frac{1}{\tau_B}}{\left(-i\omega+\frac{1}{\tau_B}\right)^3};
\label{corrdbb}\\
\Delta{\cal D}_{jB} =
\Delta{\cal D}_{Bj}= -\frac{\tau_B}{\tau_j}
 \Delta{\cal D}_{BB}, \quad j=L,R;\\
\label{corrdbl}
\Delta{\cal D}_{LR} =
\frac{1-2i\omega{\tau_B}}{\omega^2\tau_L\tau_R}
 \Delta{\cal D}_{BB}\left(\omega\right).
\label{corrdlr}
\end{eqnarray} Corrections $\Delta{\cal D}_{jj}$ for $j=L,R$ are found
with help of Eq.~(\ref{chconserv}) and they are given by
\begin{equation}
\Delta{\cal D}_{jj} =\left(
\frac{2i\omega{\tau_B}-1}{\omega^2\tau_L\tau_R}
 + \frac{\tau_B}{\tau_j}\right) 
\Delta{\cal D}_{BB}.
\label{corrdll}
\end{equation}
\end{mathletters} 

Substituting Eqs.~(\ref{corrdbb}) and (\ref{corrdbl}) into
Eq.~(\ref{neutrality}), we observe that the voltage in the cavity $V_B$
does not acquire any quantum corrections, $V_B =
Vg_L/\left(g_R+g_L\right)$.  Finally, substituting Eqs.~(\ref{corr10})
into Eq.~(\ref{gd}) and using Eq.~(\ref{taug}), we obtain the final result
for the frequency dependent weak localization correction to the 
conductance of the ballistic cavity
\begin{equation}
\Delta g(\omega) = -\frac{se^2}{2\pi\hbar}
\frac{g_Lg_R}{\left(g_L+g_R\right)^2}
\left[\frac{\Gamma(\omega)}{1-i\omega
\tau_B}\right],
\label{bmainresult}
\end{equation} where the total escape time $\tau_B$ is defined in
Eq.~(\ref{Dbbclassical})
 We emphasize that   Eq.~(\ref{bmainresult}) at zero frequency can be
obtained without electroneutrality requirement.

Equation (\ref{bmainresult}) is the main result of this section. At
$\omega =0$, this result  agrees with the findings of
Ref.~[\onlinecite{Mello}] in the  limit of large number of quantum
channels in the contact. We are  aware of neither any calculation at
finite frequency nor of a description of the role of the Ehrenfest time
in the conductance of the ballistic cavities. The renormalization function
$\Gamma(\omega)$ in Eqs.~(\ref{bmainresult}) describes the effect of the
Lyapunov region on the weak localization and it is given by
Eqs.~(\ref{result}) and (\ref{tE}).  Analytic calculation of the Lyapunov
exponents
$\lambda, \lambda_2 \simeq \tau_{fl}^{-1}$ for the ballistic cavity is a 
separate problem and it will not be done in this paper. It is assumed in
Eq.~(\ref{bmainresult}), that the condition $t_E \ll \tau_B$ holds. The
result for the opposite limit, (which corresponds to the exponentially
small Planck constant),  is obtained by substitution 
$\Gamma(\omega) \to \Gamma(\omega + i/\tau_B)$ in
Eq.~(\ref{bmainresult}) and the weak
localization correction turns out to be suppressed by the factor
$\exp\left(-2t_E/\tau_B\right)$.

The finite phase relaxation time
$\tau_\varphi$ is taken into account by substitution $\omega \to \omega
+i/\tau_\varphi$ in Eq.~(\ref{cbresult}).  At $\tau_\varphi\gg t_E$, the
result for $dc$-  conductance agrees with the result of
Ref.~[\onlinecite{Mellophase}].  We obtained for
$\omega =0$
\begin{eqnarray}
&&\Delta g =
-\frac{se^2}{2\pi\hbar}\frac{{g_Lg_R}}{\left(g_L+g_R\right)^2}
\frac{\tau_i}{\tau_B}\times \label{btauphi}\\
&&\hspace{2cm}\exp\!\left[-\frac{t_E}{\tau_i}\!
\left( 1\!-\!\frac{\lambda_2}{\lambda^2{\tau_i}}\right)
-\frac{t_E}{\tau_B}\!
\left( 1\!-\!\frac{\lambda_2}{\lambda^2{\tau_B}}
\!\right)\right]
,
\nonumber
\end{eqnarray}
where  $\tau_i$ is the time it takes for an electron
to be scattered inelastically or to escape the cavity, 
\[
\frac{1}{\tau_i}=\frac{1}{\tau_B}+\frac{1}{\tau_i}.
\]
 Usually, the Ehrenfest time $t_E$ is much smaller than the escape time
$\tau_B$. In this case, one can immediately see the dramatic crossover at
the temperature dependence (usually $\tau_\varphi$ is a power function of
temperature, see Ref.~[\onlinecite{AronovAltshuler}]). If at
$\tau_\varphi \gg t_E$,
 the dependence on temperature is a power law, with the increase of the
temperature the
 change to the exponential drop occurs. Thus, the study of the crossover
in the temperature or frequency  dependence of the ballistic cavities may
provide the information about the values and the distribution of the
Lyapunov  exponents in the cavity. 

\section{Conclusion} In this paper we developed a theory for the weak
localization (WL) correction in a quantum chaotic system, i.e. in a
system with the characteristic spatial scale of the static potential,
$a$, being much larger than the Fermi wavelength,
$\lambda_F$. We showed that for the quantum chaos, new frequency domain
appears, $t_{E}^{-1}\ll \omega \ll\tau_{tr}^{-1}$, [$t_E$ is the Ehrenfest
time, see Eq.~(\ref{tE})] where the classical  dynamics is still governed
by the diffusion equation, but the WL correction deviates from the
universal law. For the first time, we were able to investigate frequency
dependence of the WL correction at such frequencies, see
Eqs.~(\ref{renfunction}) and (\ref{result}), and to find out how the
fundamental characteristic of the classical chaos appears in the quantum
correction. At lower frequencies,
$\omega \ll t_E^{-1}$, we proved the universality of the weak localization
correction for the disorder potential of an arbitrary strength and spatial
size.

These  results may be experimentally checked by studying  the  frequency
or temperature (via $\tau_\varphi$) dependence of  the weak localization
correction ({\em e.g.} negative magnetoresistance). Indeed, at the
low-frequency or temperature, the conventional dependence should be
observed. This dependence is rather weak (logarithmical for large samples
and a power law for the ballistic cavities). With the increase of the
frequency or temperature, the dependence becomes  exponential; such a
crossover may be used to find the Ehrenfest time, $t_E$ and thus extract
the value of the Lyapunov exponent. The parameters of the ballistic
cavities studied in Ref.~[\onlinecite{Chang}] are $a\simeq 1\mu m,\
\lambda_F \simeq 400
\AA$, so that $\ln(a/\lambda_F) \simeq 3$. We believe, however, that the
size of the ballistic cavities may be raised up to the mean free path 
$\simeq 17\mu m$; Ehrenfest time in this case would be appreciably larger
than the flying time, $\ln(a/\lambda_F) \simeq 6$, and the characteristic
frequency $\omega = t_E^{-1}$ for this case can be estimated as
$\omega \simeq 5\times 10^9 s^{-1}$. Measurements of frequency dependence
of the WL correction in quantum disorder regime were performed in
Ref.~[\onlinecite{Carini}] at frequency  as high as $16.5\ {\rm GHz}$. 
Thus, the measurement of the Ehrenfest time in the ballistic cavities
does not seem to be unrealistic. 

We expect that the effects associated with the Ehrenfest time may be found
also in optics. They may be observed, {\em e.g.} in the dependence of the
enhanced backscattering on  frequency of the amplitude modulation
$\omega$, This dependence should be still given by our function
$\Gamma(\omega)$ with
$\lambda_F$ being replaced with the light wavelength.

We showed that the description of the intermediate region $t_E^{-1} \ll
\omega\ll
\tau_{tr}^{-1}$ can be reduced to the solution of the purely classical
equation of motion, however, the averaging leading to the Boltzmann
equation is not possible because the initial and final phase cells of the
relevant classical correlator (Cooperon) are related by the time
inversion. Therefore, the initial and finite segments of the
corresponding classical trajectory are strongly correlated and their
relative motion is described by the Lyapunov exponent and not by the
diffusion equation. We took this correlation into account, showed that it
is described by the log-normal  distribution function  and  related the
Ehrenfest time to the parameters of this function.

Because the description by the Boltzmann equation was not possible, we
derived  the lowest order quantum correction to the classical correlator
in terms of the  solution of the Liouville equation, smeared by the small
angle diffraction, see Eq.~(\ref{mainresult0}).  The derivation was based
on the equations of motion for the exact Green functions and did not
imply  averaging over the realization of the potential.

Closing the paper, we would like to discuss its relation to the other
works and to make few remarks concerning how the Ehrenfest time appears
in the level statistics.

First, we notice that, though quite popular in the classical mechanics and
hydrodynamics, the Lyapunov exponent very rarely enters in the expressions
for  observable quantities in the solid state physics, see
Ref.~[\onlinecite{Larkin68}]. The possibility to observe the intermediate
frequency region $t_E^{-1} \ll \omega
\ll \tau_{tr}$ appeared only recently with the technological advances in
the preparing of the ballistic cavities and that is why the region has
not been studied systematically as of yet. Let us mention that the
importance of the Ehrenfest time in the semiclassical approximation was
noticed already in Ref.~[\onlinecite{Larkin68}] where it was shown that
the method of quasiclassical trajectories in the theory of
superconductivity\cite{diGennesShapoval} fails to describe some
non-trivial effect at times larger than $t_E$ which was calculated for
the dilute scatterers. The term  ``Ehrenfest time'' for the quantity
(\ref{tE}) was first introduced in Ref.~[\onlinecite{Shepelyanskii}]
 The relevance of $t_E$ in the theory of weak localization was emphasized
by Argaman\cite{Argaman}, however, he focused only on times much larger
than the Ehrenfest time.

The universality of weak localization correction at small frequencies was
known for the case of weak quantum impurities\cite{Gorkov} and for the
ballistic cavities\cite{Mello}. We are not aware of any proof of the
universality for the disorder potential of arbitrary strength and spatial
scale.

The description of the quantum corrections in terms of the nonaveraged
solutions of the Liouville equation was developed in by Muzykantskii and
Khmelnitskii\cite{Muzykantskii} and more recently by Andreev {\em et.
al.}\cite{Andreev2}, who suggested  the effective
supresymmetric\cite{efetovadvances} action in the ballistic regime.  In
Ref.[\onlinecite{Andreev2}], the supersymmetric action was written in
terms of the Perron-Frobenius operator which differs from the first order
Liouville operator by the regularizator of the second order. This
regularizator is similar to the angular diffusion term, $\propto
1/\tau_q$ in Eqs.~(\ref{Mainresult}). The authors mentioned that all the
physical results can be obtained if the limit of vanishing regularizator
is taken in the very end of the calculation. Our finding indicate that
the time it takes for the quantum correction to reach its universal value
is
$\propto \ln\left(\tau_q\right)$. Thus, at any finite frequency, the limit
$\tau_q
\to\infty$ can not be taken and the regularizator in the supersymmetric
action should be assigned its physical value, see Eq.~(\ref{tauestem}). 

 In principle, our formula for the weak localization correction
(\ref{mainresult0}) can be derived using the supersymmetry technique.
However, our approach seems to be technically easier and more physically
tractable for the calculation of the first order weak localization
correction. We believe that the supersymmetry may serve as a powerful 
tool for the investigation of the effect of the Ehrenfest time on the
higher order corrections and on the level statistics.

It is generally accepted that the level statistics at low energies is
described by the Wigner-Dyson distribution\cite{book}. For the small
disordered particle it was first proven by Efetov\cite{efetovadvances} and
for the ballistic cavities by   Andreev {\em et. al.}\cite{Andreev2}. For
the quantum disorder, Altshuler and Shklovskii\cite{AltshulerShklovskii}
showed that the universal Wigner-Dyson statistics breaks down at the
Thouless energy.
 For the ballistic cavities the universal statistics is believed to be
valid up to the energies of the order of the inverse flying time
$\tau_{fl}$, at smaller energies $s$ the corresponding corrections are
small as $s\tau_{fl}$.  We, however, anticipate deviations at the
parametrically smaller energies of the  order of $t_E^{-1}$, and the
corrections  of the order of
$st_E$ at energies $s\ll t_E^{-1}$.  

Let us consider for concreteness the correlator of the density of states
${\cal R}(s) = \langle\rho(\epsilon)\rho(\epsilon+s)\rangle -
\langle\rho(\epsilon)\rangle^2$, where
$\rho(\epsilon)=Tr\delta\left(\epsilon-\hat{H}\right)$. For the orthogonal
gaussian ensemble the random matrix theory yields\cite{book}
${\cal R}(s) = -\left(\pi s\right)^{-2}+ 
\left(1+\cos^2\pi s\right)/\left(\pi s\right)^{-4} +\dots$, where $s\gg
1$ is measured in units of mean level spacing. We expect, that the first
term in this expression is not affected by the presence of the Lyapunov
region, whereas the following terms are. In the supersymmetric
approach\cite{efetovadvances} this  follows from the fact that the first
term arises from non interacting diffuson modes whereas all the others
come from the interaction of these modes.  Such interaction is analogous
to the one giving rise to the weak localization which was shown to have
the frequency dispersion described by the renormalization function
$\Gamma (\omega)$, see Eq.~(\ref{result}). We believe that the same
renormalization factor will appear in all the effects associated with the
coupling of the diffuson-Cooperon modes.

\acknowledgements We are thankful to V.I. Falko for discussions of the
results, to  H.U. Baranger and L.I. Glazman for reading the manuscript
and valuable remarks, and to D.L. Shepelyansky for pointing
Ref.~[\onlinecite{Shepelyanskii}] to us. One of us (I.A.) was supported
by NSF Grant DMR-9423244.
\appendix
\section{Lyapunov exponent for the weak scatterers}
\label{ap:1} We consider explicitly the case where the potential $U$ in
Eq.~(\ref{wequationt}) is weak and its distribution function is Gaussian.
For the sake of simplicity we neglect the angular diffusion due to the
quantum impurities in the Lyapunov region, because this diffusion does not
affect  values of $\lambda,\ \lambda_2$, see Sec.~\ref{sec:4.4}. In this
case it is more convenient not to follow the general procedure outlined in
Sec.~(\ref{sec:4.4}), but to make use of the small parameter
$a/\l_{tr}$ first. Considering the disorder potential in the second order
of the perturbation theory, we obtain for independent on ${\bf R}, {\bf
n}$ part of the function
$W_\perp$:
\begin{equation}
\left[
\frac{\partial}{\partial t} - v_F\phi\frac{\partial}{\partial \rho}
-\frac{2}{\tau_{tr}}{\cal E}(\rho)\frac{\partial^2}{\partial\phi^2}
\right] W_\perp =0,
\label{corkin}
\end{equation} where the transport life time is given by
\begin{equation}
\frac{1}{\tau_{tr}} = \frac{1}{4\epsilon_Fp_F}\int_{-\infty}^\infty dx 
\langle \partial_yU(x,0)\partial_yU(0,0)\rangle,
\label{tlt}
\end{equation} and the dimensionless function ${\cal E}$ is defined as 
\begin{equation} {\cal E}(\rho) = 1- \frac{\int_{-\infty}^\infty dx 
\langle \partial_yU(x,\rho)\partial_yU(0,0)\rangle} {\int_{-\infty}^\infty
dx 
\langle \partial_yU(x,0)\partial_yU(0,0)\rangle}.
\label{E}
\end{equation} In  Eq.~(\ref{corkin}), we assumed $\phi \ll 1$ only and
lifted the other assumption of Eq.~(\ref{wequationt}) $\rho \ll a$.
 If $\rho \ll a$, we expand ${\cal E}$ in Taylor series, ${\cal E}(\rho)
\approx
 \rho^2/2a^2$, which rigorously defines length $a$ in this case, and we
arrive to the equation describing the Lyapunov region for  the weak
disorder potential 
\begin{equation}
\left[
\frac{\partial}{\partial t} - v_F\phi\frac{\partial}{\partial \rho}
-\frac{\rho^2}{\tau_{tr}a^2}\frac{\partial^2}{\partial\phi^2}
\right] W_\perp =0.
\label{corkinL}
\end{equation} It is worth noticing, that our approach is equivalent to
one involving the multiplication of the vector $(\rho,\phi)$ by a
Monodromy matrix after each scattering event.
 Equation (\ref{corkinL}) is valid because each Monodromy matrix defined
on a time of the order $a/v_F$ is close to unit matrix. Otherwise, the
last term in the brackets in Eq.~(\ref{corkinL}) becomes an integral
operator.

After introduction of new variables 
\begin{equation} z = \ln \frac{a}{\rho}, \quad y=
\frac{a\phi}{\rho}\left(\frac{l_{tr}}{a}\right)^{1/3}\!\!\!, \quad
\tau = \frac{t}{\tau_{tr}}\left(\frac{l_{tr}}{a}\right)^{2/3}\!\!\!,
\label{nva}
\end{equation} equation (\ref{corkinL}) acquires a simple form
\begin{equation}
\left[
\frac{\partial}{\partial \tau} - y
\frac{\partial}{\partial z} + y^2\frac{\partial}{\partial y} -
\frac{\partial^2}{\partial y^2}
\right] W_\perp =0.
\label{simpleform}
\end{equation}

We are interested in the case when function $W_\perp$ changes slowly as
the function of $z$. Corresponding gradient is small, and we can employ
the procedure similar to the reducing the Boltzmann equation to  the
diffusion equation. Let us represent function
$W_\perp$ as
\begin{equation} W_\perp(\tau; z, y) =W_\perp^0(\tau; z) + W^1(\tau;
z,y), \quad  W_1\ll W_\perp.
\label{representation}
\end{equation} Substituting Eq.~(\ref{representation}) into
Eq.~(\ref{simpleform}),
 multiplying the result by function $g(y)$:
\begin{equation}
\left[\frac{d}{d y}y^2+\frac{d^2}{d y^2}\right]g(y)=0, \quad \int dy g(y)
=1.
\label{g}
\end{equation} and integrating over $y$ we obtain
\begin{equation}
\left[
\frac{\partial}{\partial \tau} - \beta
\frac{\partial}{\partial z}\right] W_\perp^0 -
\frac{\partial}{\partial z}\!\int\!d y \left[\left(y-\beta\right)
 g(y) W^1(y)\right]  =0,
\label{multipliedeq}
\end{equation} where  the numerical coefficient $\beta$ is given by
\begin{equation}
\beta = \int dy \ y g(y)
\label{drift}
\end{equation} and function $W^1$ can be written as
\begin{equation} W^1 = h(y) \frac{\partial W_\perp^0(\tau;z)}{\partial
z}, \ 
\left[y^2\frac{d}{d y}-\frac{d^2}{d y^2}\right]h(y)= y-\beta. 
\label{h}
\end{equation} Shift of $W^1$ by an arbitrary constant does not affect
the results,  see Eqs.~(\ref{diff}) and (\ref{drift}). Substituting
Eq.~(\ref{h}) into Eq.~(\ref{multipliedeq}), we obtain in  accordance
with general formula (\ref{difflog})
\begin{equation}
\left[
\frac{\partial}{\partial \tau} - \beta
\frac{\partial}{\partial z}  - \beta_2
\frac{\partial^2}{\partial z^2}
\right] W_\perp^0(\tau;z) =0,
\label{diffloga}
\end{equation} where the numerical coefficient $\beta_2$ is given by
\begin{equation}
\beta_2 = \int dy \ \left(y-\beta\right) g(y)h(y).
\label{diff}
\end{equation} Comparing Eqs.~(\ref{diffloga}) and (\ref{nva}) with
Eq.~(\ref{difflog}),  we find for Lyapunov exponent $\lambda$ and its
deviation $\lambda_2$ 
\begin{equation}
\lambda = \frac{\beta}{\tau_{tr}} \left(\frac{l_{tr}}{a}\right)^{2/3},
\quad
\lambda_2 = \frac{2 \beta_2}{\tau_{tr}}
\left(\frac{l_{tr}}{a}\right)^{2/3}.
\end{equation} Simple calculation of the numeral coefficients $\beta,
\beta_2$ is
 carried out with the help of Eqs.~(\ref{drift}), (\ref{g}), (\ref{diff})
and (\ref{h})  with the final result
\begin{eqnarray} &&\displaystyle{\beta = \frac{\int_{-\infty}^{\infty}
dy\ e^{-y^3/3} y
\int_{-\infty}^y dy_1\ e^{y^3_1/3}} {\int_{-\infty}^{\infty} dy\
e^{-y^3/3} 
\int_{-\infty}^y dy_1\ e^{y^3_1/3}} \approx 0.365,  }
\label{resbeta}\\ &&\displaystyle{\beta_2 = 
\frac{
         \int_{-\infty}^{\infty}\!\! dy
         \left[
              e^{y^3/3}\!\!\int_{-\infty}^y\!\!dy_1 e^{y^3_1/3}\!\!
               \left(
                     \int^{\infty}_y\!dy_2 (\beta\!-\! y_2) e^{-y^3_2/3}
                \right)^2
          \right] } {
\int_{-\infty}^{\infty} dy\ e^{-y^3/3} 
\int_{-\infty}^y dy_1\ e^{y^3_1/3} }  }\nonumber\\
&&\displaystyle{\hspace{5cm}\approx 0.705.}\nonumber
\end{eqnarray} In order to avoid any confusion, let us notice that the
log-normal distribution function can not be used to find  the averaged
moments of the coordinates
$\rho,\phi$ and it is sufficient only for the calculation of the low
moments of the logarithm of the coordinates.

\end{multicols}

\begin{references}
\bibitem{footnote1} This result, strictly speaking, is applicable for the
case of the strong scatterers where the transport and elastic mean free
paths are of the same order. For the case of weak scatterers one should
use
$v_F/\lambda$ instead of
$l_{tr}$. Here $\lambda$ is the Lyapunov exponent, see Appendix
~\ref{ap:1}.
\bibitem{interaction}The interaction effects analogous to those in the
theory of QD\cite{AronovAltshuler} are beyond the scope of our paper.
\bibitem{footnote2}Actually, these arguments give 
 estimate $t_E =
\frac{1}{\lambda}\ln\left(\frac{a^2}{l_{tr}\lambda_F}\right)$ rather than
Eq.~(\ref{tE}), however, the corresponding error in the definition of the
Ehrenfest time $\delta t_E \simeq
\frac{1}{\lambda}\ln\left(\frac{l_{tr}}{a}\right)$ is of the order of the
$\tau_{tr}$. Because all the procedure is valid only for
$\omega\tau_{tr}\ll 1$, keeping such terms in the exponent would be
beyond the accuracy of the calculation.
\end{references}
\end{document}